\documentclass[a4paper,fleqn,usenatbib]{mnras}

\usepackage[T1]{fontenc}

\usepackage[english]{babel}
\usepackage{graphicx}
\usepackage{fleqn}
\usepackage{amssymb}
\usepackage{amsmath}
\usepackage{booktabs}
\usepackage{hyperref}
\usepackage{txfonts} 
\usepackage{comment}
\usepackage{color, colortbl}

\newcommand{\Se}{Section}
\newcommand{\F}{Fig.}

\newcommand{\msun}{\mathrm{M}_\odot}

\newcommand{\au}{\,\textsc{au}}

\newcommand{\bin}{\mathrm{bin}}
\newcommand{\p}{\mathrm{p}}
\newcommand{\m}{\mathrm{m}}

\renewcommand{\pm}{\mathrm{pm}}
\newcommand{\MMC}{\mathrm{MMC}}

\newcommand{\peri}{\mathrm{peri}}
\newcommand{\crit}{\mathrm{crit}}

\title[Moons around circumbinary planets]{Stability of exomoons around the {\it Kepler} transiting circumbinary planets}

\author[Hamers et al.]{Adrian S. Hamers$^{1}$\thanks{E-mail: hamers@ias.edu}, Maxwell X. Cai$^{2}$\thanks{cai@strw.leidenuniv.nl}, Javier Roa$^{3}$\thanks{javier.roa@jpl.nasa.gov}, and Nathan Leigh$^{4,5}$\thanks{nleigh@amnh.org} \\
$^{1}$Institute for Advanced Study, School of Natural Sciences, Einstein Drive, Princeton, NJ 08540, USA \\
$^{2}$Leiden Observatory, Leiden University, PO Box 9513, NL-2300 RA Leiden, The Netherlands \\
$^{3}$Jet Propulsion Laboratory, California Institute of Technology, 4800 Oak Grove Dr, Pasadena, CA 91109, USA \\
$^{4}$Department of Astrophysics, American Museum of Natural History, Central Park West and 79th Street, New York, NY 10024, USA \\
$^{5}$Department of Physics and Astronomy, Stony Brook University, Stony Brook, NY 11794-3800, USA}
\date{Accepted 2018 August 01. Received 2018 August 01; in original form 2018 June 15.}

\pubyear{2018}

\begin{document}
\label{firstpage}
\pagerange{\pageref{firstpage}--\pageref{lastpage}}
\maketitle

\begin{abstract} 
The {\it Kepler} mission has detected a number of transiting circumbinary planets (CBPs). Although currently not detected, exomoons could be orbiting some of these CBPs, and they might be suitable for harboring life. A necessary condition for the existence of such exomoons is their long-term dynamical stability. Here, we investigate the stability of exomoons around the {\it Kepler} CBPs using numerical $N$-body integrations. We determine regions of stability and obtain stability maps in the $(a_\m,i_\pm)$ plane, where $a_\m$ is the initial exolunar semimajor axis with respect to the CBP, and $i_\pm$ is the initial inclination of the orbit of the exomoon around the planet with respect to the orbit of the planet around the stellar binary. Ignoring any dependence on $i_\pm$, for most {\it Kepler} CBPs the stability regions are well described by the location of the 1:1 mean motion commensurability of the binary orbit with the orbit of the moon around the CBP. This is related to a destabilizing effect of the binary compared to the case if the binary were replaced by a single body, and which is borne out by corresponding 3-body integrations. For high inclinations, the evolution is dominated by Lidov-Kozai oscillations, which can bring moons in dynamically stable orbits to close proximity within the CBP, triggering  strong interactions such as tidal evolution, tidal disruption, or direct collisions. This suggests that there is a dearth of highly-inclined exomoons around the {\it Kepler} CBPs, whereas coplanar exomoons are dynamically allowed.
\end{abstract}


\section{Introduction}
\label{sect:introduction}
One of the exotic type of planetary systems detected by the {\it Kepler} mission are transiting circumbinary planets (CBPs), i.e., planets in nearly-coplanar orbits around a stellar binary (and nearly coplanar with the plane of the sky), temporarily blocking the binary's light and giving a generally complex light curve (e.g., \citealt{2015MNRAS.449..781M,2017MNRAS.465.3235M}). Currently, 10 confirmed {\it Kepler} CBPs are known in 9 binary systems (see Table\,\ref{table:IC} for an overview). These systems have revealed important clues for the formation and evolution of planets and high-order multiple systems. For example, none of the {\it Kepler} CBPs are orbiting binaries with periods shorter than 7 d, which suggests the presence of a third star \citep{2015PNAS..112.9264M,2015MNRAS.453.3554M,2016MNRAS.455.3180H}, or could be indicative of coupled stellar-tidal evolution \citep{2018ApJ...858...86F}. 

A key question for the CBP systems is whether they would be able to harbour life. As shown by various authors \citep{2012Sci...337.1511O,2012ApJ...750...14Q,2013ApJ...762....7K,2013ApJ...777..166H,2014ApJ...780...14C,2015ApJ...798..101C,2016ApJ...818..160Z,2017AJ....154..157W,2018arXiv180206856M}, some of the {\it Kepler} CBPs are within the habitable zone (HZ). Some of the {\it Kepler} CBPs are giant planets and are unlikely to be able to sustain life; on the other hand, exomoons are prime candidates for celestial objects harbouring life (e.g., \citealt{1987AdSpR...7..125R,1997Natur.385..234W,2013AsBio..13...18H,2014OLEB...44..239L,2017AN....338..413S,2017A&A...601A..91D}). Although techniques to observe exomoons have been developed, they have not yet been found (e.g., \citealt{2009MNRAS.392..181K,2009MNRAS.396.1797K,2012ApJ...750..115K,2013ApJ...770..101K,2013ApJ...777..134K,2014ApJ...784...28K}; see \citealt{2014arXiv1405.1455K,2014AsBio..14..798H,2017haex.bookE..35H} for reviews). Given the abundance of moons in the Solar system, it is reasonable to assume that they exist. 

A necessary condition for the existence of exomoons in the {\it Kepler} CBPs is that they be long-term dynamically stable. In this paper, we address this issue by carrying out $N$-body integrations of exomoons orbiting around the {\it Kepler} CBPs (i.e., in S-type orbits around the CBPs, \citealt{1984CeMec..34..369D}). We will show that stable configurations are possible for a wide range of parameters, and that the stability boundary is well described by the location of the 1:1 mean motion commensurability with the stellar binary. 

Previous theoretical efforts have focused on the inclined Hill stability of test particles around single stars (e.g., \citealt{1979AJ.....84..960I,1980AJ.....85...81I,1991Icar...92..118H,2003AJ....126..398N,2017MNRAS.466..276G}), or on stability from a secular (i.e., orbit-averaged) point of view in the case of hierarchical quadruple systems (e.g., \citealt{2017ApJ...836...27M,2017MNRAS.470.1657H,2018MNRAS.474.3547G}). Other studies have revealed a number of interesting properties that distinguish CBPs from single-star host systems. For example, \citet{smullen16} showed using $N$-body simulations that the disruption of multi-planet systems tends to result in far more ejections relative to single-host systems, which more commonly lose their planets due to collisions upon becoming dynamically unstable. Typically, numerical simulations suggest of order 80\% of outcomes correspond to ejections, and only 20\% to physical collisions with the primary or secondary \citep{sutherland16}. In this chaotic regime, the distribution of escaper velocities has been well-studied and parameterized in previous works, mostly in the context of scattering of single stars by super-massive black hole binaries \citep[e.g.,][]{quinlan96,sesana06}.  In addition, planet-planet scattering in systems with CBPs can lead to S-type tidal capture planets in close binaries (e.g., \citealt{2018arXiv180505868G}).

A previous work that focussed directly on the stability of exomoons around CBPs is \citet{2012ApJ...750...14Q}, who considered the dynamical stability (and habitability) of exomoons around Kepler 16. However, \citet{2012ApJ...750...14Q} considered several orbital configurations of the exomoon, and did not focus in detail on S-type moons around the CBP. Also, the effect of the mutual inclination of the orbit of the moon with respect to the orbit of the CBP around the stellar binary was not considered. Here, we carry out a more detailed and systematic study of Hill-stable S-type orbits around CBPs, considering all possible inclinations, and including all currently-confirmed {\it Kepler} CBP systems.

The structure of this paper is as follows. In \Se\,\ref{sect:methodology}, we briefly describe the initial conditions and the numerical methodology used for the $N$-body integrations and for determining stability regions. The main results are given in \Se\,\ref{sect:results}, in which we show the stability maps for the {\it Kepler} systems. We discuss our results in \Se\,\ref{sect:discussion}, and we conclude in \Se\,\ref{sect:conclusions}.

\section{Methodology}
\label{sect:methodology}

\begin{figure}
\center
\includegraphics[scale = 0.6, trim = 10mm 20mm 0mm 0mm]{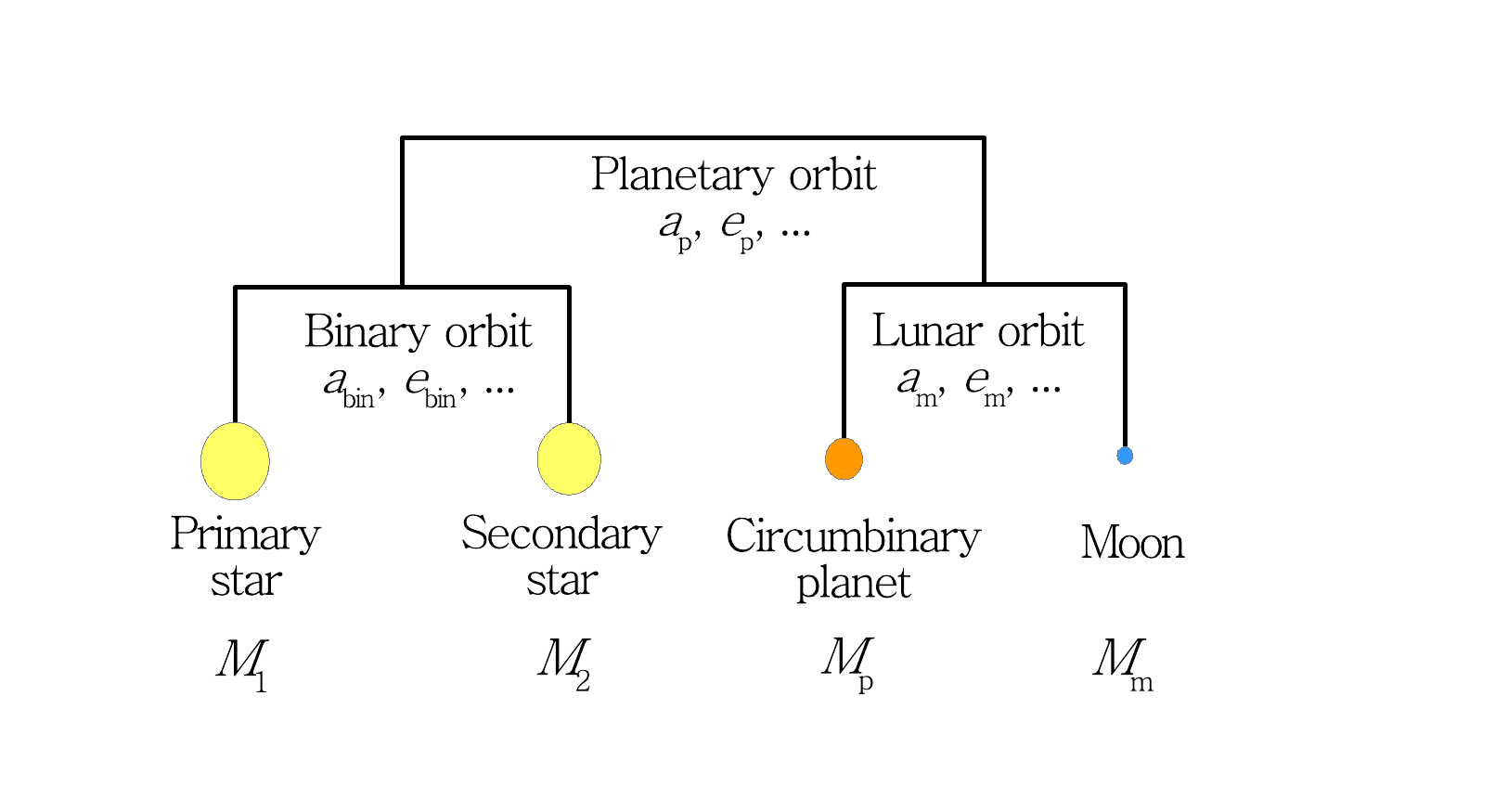}
\caption{\small Schematic representation (not to scale) of the orbits of the stellar binary, the CBP, and its moon. }
\label{fig:configuration}
\end{figure}

\subsection{Initial conditions}
\label{sect:methodology:IC}
Our systems consist of a stellar binary (masses $M_1$ and $M_2$; semimajor axis $a_\bin$) orbited by a CBP (mass $M_\p$ and radius $R_\p$; semimajor axis $a_\p$). In turn, the CBP is orbited by a moon (mass $M_\m$; semimajor axis $a_\m$). A schematic representation (not to scale) in a mobile diagram of the system \citep{1968QJRAS...9..388E} is given in \F\,\ref{fig:configuration}. 

We take the binary and planetary parameters for the currently-known {\it Kepler} systems from various sources. The adopted orbital parameters are given for convenience and completeness in Table\,\ref{table:IC}. In most cases, the orbital angles are defined with respect to the plane of the sky. Note that the binary's longitude of ascending node, $\Omega_\bin$, is defined to be zero. For several systems, the argument of periapsis of the binary or the planet is not known. In this case, we set the corresponding values to zero. We do not vary the initial mean anomalies of the binary and the planet (both are set to zero), but we do vary the moon's initial mean anomaly (see below). 

For each binary+planet system, we set up a fourth particle, the exomoon (henceforth simply `moon'), with mass $M_\m = q_\m M_\p$, where $q_\m$ is either 0.001 or 0.1, initially orbiting around the CBP (for reference, $q_\m \simeq 0.012$ for the Earth-Moon system, $q_\m \simeq 4.7 \times 10^{-5}$ for Io-Jupiter, and $q_\m \simeq 0.12$ for Pluto-Charon). We consider 40 values of the lunar semimajor axis, $a_\m$, on an evenly-spaced grid, with the lower and upper values depending on the system; typically, $a_\m$ ranges between $0.005$ and $0.02\,\au$ (see the figures in \Se\,\ref{sect:results:stab} for the precise ranges of $a_\m$ assumed for each system). The initial eccentricity is set to $e_\mathrm{m}=0.01$; we also carried out simulations with higher initial eccentricities, but found no significant dependence of the stability regions on $e_\mathrm{m}$. 

The lunar argument of periapsis is set to $\omega_\m = 0$; the longitude of the ascending node is set to $\Omega_\m = \Omega_\p + \pi$, such that the mutual inclination between the planet and the moon is
\begin{align}
\label{eq:i_pm}
\nonumber \cos(i_\pm) &= \cos(i_\p) \cos(i_\m) +  \sin(i_\p) \sin(i_\m) \cos(\Omega_\p - \Omega_\m) \\
&= \cos(i_\p + i_\m),
\end{align}
i.e., by construction, the mutual inclination between the planet and the moon is $i_\pm = i_\p + i_\m$. The inclination of the moon is set to $i_\m = i_\mathrm{grid} - i_\p$, where $i_\mathrm{grid}$ is evenly spaced from 0 to $180^\circ$ with 40 values. We consider 10 different values of the lunar initial mean anomaly, $\mathcal{M}_\m$, evenly spaced between 0 and $2\pi$. In summary, for each {\it Kepler} system in Table\,\ref{table:IC}, we carry out $40\times40\times10 = 16,000$ integrations with different $a_\m$, $i_\pm$ and $\mathcal{M}_\m$. These sets are repeated for $q_\m = 0.001$ and $q_\m = 0.1$ (binary case), and for $q_\m=0.001$ and with the stellar binary effectively replaced by a point mass (single case, i.e., the primary stellar mass is replaced by $M_1+M_2$, and the secondary stellar mass is reduced to $M_2=1\,\mathrm{kg}$, essentially a test particle). This implies a total of $10\times3\times16,000=480,000$ simulations for all 10 {\it Kepler} CBP systems. 

We carry out 4-body integrations of the binary+CBP+moon system for a duration of $t_\mathrm{end} = 1000\,P_\p$, where $P_\p$ is the orbital period of the planet around the binary. To investigate whether this integration time is sufficiently long, we show the stability regions below in \Se\,\ref{sect:results} for different integration lengths. The (absolute values of the) relative energy errors in the simulations after $1000\,P_\p$ are typically on the order of $10^{-12}$, with maxima of $\sim 10^{-10}$.

\subsection{Numerical integrations}
\label{sect:methodology:num}
For our integrations, we use \textsc{ABIE}, which is a new \textsc{Python}-based code with part of the code written in the \textsc{C} language for optimal performance (Cai et al., in prep.). \textsc{ABIE} includes a number of integration schemes. Here, we use the Gau\ss-Radau scheme of 15th order \citep{everhart1985efficient}, with a minimum time-step of $10^{-13}$, and a tolerance parameter of $\epsilon = 10^{-8}$. The integrator incorporates a more sophisticated step-size control algorithm proposed by \citet{2015MNRAS.446.1424R}. The scheme is optimized for handling close encounters.

We only include purely Newtonian dynamics in our integrations. Strong interactions and short-range forces such as tidal effects, tidal disruption, and the oblateness of the stars and planet due to rotation are considered posteriori by considering the periapsis distances in the simulations, in \Se\,\ref{sect:results:int}. 

We do check for collisions of the moon with the planet or the stars, taking the collision radii to be the observed physical radii. In reality, the moon is likely tidally disrupted before colliding with the CBP or with the stars. Generally, the tidal disruption radius depends on the density of the moon, i.e., in the case of tidal disruption by the planet, \citep{roche1847}
\begin{align}
r_\mathrm{Roche} = 1.523 \left ( \frac{M_\p}{\rho_\m} \right )^{1/3} \simeq 1.8\,R_\mathrm{J} \,\left (\frac{M_\p}{1\,M_\mathrm{J}} \right )^{1/3} \left ( \frac{\rho_\m}{3.34\,\mathrm{g\,cm^{-3}}} \right )^{-1/3},
\end{align}
where $\rho_\m$ is the density of the moon. For the numerical estimate, we set $M_\p=M_\mathrm{J}$ (disruption by a Jupiter-mass planet) and $\rho_\m=3.34\,\mathrm{g\,cm^{-3}}$, appropriate for the Moon. However, for CBPs with relatively large radii and for high density moons, $r_\mathrm{Roche}$ could be less than $R_\p$. Given that the lunar density is unknown, we here choose a more general approach of setting the collision radii equal to the physical radii. We note that larger collision radii would lead to a larger number of collisions, in particular with the CBP (see \Se\,\ref{sect:results:int} below).

\subsection{Determining stability regions}
\label{sect:methodology:regions}
We determine stability maps from our 4-body integrations using the following approach. For a given integration time $t_\mathrm{end}$ (by default, $t_\mathrm{end}=1000\,P_\p$ unless specified otherwise) and a point in the $(a_\m, i_\pm)$ parameter space, we define the orbit as `stable' if the moon remains bound to the CBP (i.e., $a_\m>0$ and $0\leq e_\m < 1$) at $t_\mathrm{end}$ for all 10 realizations with different $\mathcal{M}_\m$ (green). If none of the 10 realizations of $\mathcal{M}_\m$ yield bound orbits to the CBP, then the orbit is flagged as `unstable' (red). If the orbit remains bound for some, but not all realizations with different $\mathcal{M}_\m$, then we consider the orbit to be `marginally stable' (yellow). 

\begin{table*}
\scriptsize
\begin{tabular}{lccccccccccccccc}
\toprule
Name & $M_1$ & $M_2$ & $M_\mathrm{p}$ & $R_\p$ & $a_\bin$ & $a_\p$ & $e_\bin$ & $e_\p$ & $i_\bin$ & $i_\p$ & $\Omega_\bin$ & $\Omega_\p$ & $\omega_\bin$ & $\omega_\p$ & Ref. \\
& $\msun$ & $\msun$ & $M_\mathrm{J}$ & $R_\mathrm{J}$ & $\au$ & $\au$ & & & deg & deg & deg & deg & deg & deg \\
\midrule
Kepler 16		& 0.6897		& 0.20255		& 0.333		& 0.754		& 0.22431 	& 0.7048		& 0.15944		& 0.0069		& 90.3401		& 90.0322		& 0 			& 0.003		& 263.464		& 318.0		& 1		\\
Kepler 34		& 1.0479		& 1.0208		& 0.22		& 0.764		& 0.22882		& 1.0896		& 0.52087		& 0.182		& 89.8584		& 90.355		& 0 			& -1.74		& 0			& 0			& 2		\\
Kepler 35		& 0.8877		& 0.8094		& 0.127		& 0.728		& 0.17617		& 0.60347		& 0.1421		& 0.042		& 90.4238		& 90.76		& 0 			& -1.24		& 0 			& 0 			& 2		\\
Kepler 38		& 0.949		& 0.249207	& 0.00021		& 0.388		& 0.1469		& 0.4644		& 0.1032		& 0.032		& 0			& 0.182		& 0 			& 0 			& 268.680		& 0			& 3		\\
Kepler 47b	& 1.043		& 0.362		& 0.026515	& 0.266		& 0.0836		& 0.2956		& 0.0234		& 0.035		& 89.34		& 89.59		& 0			& 0.1			& 212.3		& 0			& 4		\\
Kepler 47c	& 1.043		& 0.362		& 0.072877	& 0.411		& 0.0836		& 0.989		& 0.0234		& 0.411		& 89.34		& 89.826		& 0 			& 1.06		& 212.3		& 0			& 4		\\
Kepler 64		& 1.47		& 0.37		& 0.531558	& 0.551		& 0.1769		& 0.642		& 0.204		& 0.1			& 87.59		& 90.0		& 0 			& 0 			& 214.3		& 105.0		& 5		\\
Kepler 413	& 0.820		& 0.5423		& 0.21074		& 0.387		& 0.10148		& 0.3553		& 0.0365		& 0.1181		& 0			& 4.073		& 0 			& 0 			& 279.74		& 94.6		& 6		\\
Kepler 453	& 0.944		& 0.1951		& 0.000628	& 0.553		& 0.18539		& 0.7903		& 0.0524		& 0.0359		& 90.266		& 89.4429		& 0 			&2.103		& 263.05		& 185.1		& 7		\\
Kepler 1647	& 1.2207		& 0.9678		& 1.51918		& 1.058		& 0.1276		& 2.7205		& 0.1602		& 0.0581		& 87.9164		& 90.0972		& 0			& -2.0393		& 300.5442	& 155.0464	& 8		\\
\bottomrule
\end{tabular}
\caption{Initial conditions adopted for the {\it Kepler} CBP systems. See \F\,\ref{fig:configuration} for a schematic illustration of the definition of the orbital parameters. Here, $i$, $\Omega$ and $\omega$ denote inclination, longitude of the ascending node, and argument of periapsis, respectively. The initial mean anomalies for the binary and planetary orbit were set to $\mathcal{M}_\bin = \mathcal{M}_\p = 0$. References: 1 --- \citet{2011Sci...333.1602D}; 2 --- \citet{2012Natur.481..475W}; 3 --- \citet{2012ApJ...758...87O}; 4 --- \citet{2012Sci...337.1511O}; 5 --- \citet{2013ApJ...768..127S,2013ApJ...770...52K}; 6 --- \citet{2014ApJ...784...14K}; 7 --- \citet{2015ApJ...809...26W}; 8 --- \citet{2016ApJ...827...86K} . }
\label{table:IC}
\end{table*}

\section{Results}
\label{sect:results}
\subsection{Stability regions}
\label{sect:results:stab}
\subsubsection{General description}
\label{sect:results:stab:gen}
The main results of this paper are the stability maps shown in Figs\,\ref{fig:stab_k16} and \ref{fig:stab_krest}, for Kepler 16, and the other 9 {\it Kepler} CBPs, respectively. In these maps, each point in the $(a_\m,i_\pm)$ plane represents 10 simulations with different initial $\mathcal{M}_\m$; green filled circles correspond to stable systems, red crosses to unstable systems, and yellow open circles to marginally stable systems (see \Se\,\ref{sect:methodology:regions} for the definition of stable, unstable and marginally stable systems). If a collision of the moon with the stars or the CBP occurred for one or more $\mathcal{M}_\m$-realizations, then this is indicated with either the `$-$' or `$\star$' symbols for collisions with the planet and the stars, respectively. We show the frequency for the most common collision type among the mean anomalies, and the color of these symbols encodes this frequency with respect to the 10 realizations of $\mathcal{M}_\m$ (yellow to red for 1 to 10). For collisions of the moon with the stars, the large and small `$\star$' symbols correspond to collisions with the primary and secondary star, respectively. 

In \F\,\ref{fig:stab_k16}, the three panels in each row correspond to different integration times: $t_\mathrm{end}=400$, 700 and 1000 $P_\p$. In the first and third columns, the mass ratio of the moon to the planet is $q_\m = M_\m/M_\p = 0.001$, whereas in the second column, $q_\m = 0.1$. The third column corresponds to the single-star case, in which the stellar binary is replaced by a point mass. In \F\,\ref{fig:stab_krest}, the integration time is 1000 $P_\p$, and the mass ratio is $q_\m = 0.001$. 

As expected, the stability regions decrease for longer integration times. However, there is little dependence on the integration time for the values shown, indicating that $1000\,P_\p$ is sufficiently long to determine stability in the majority of the parameter space. In the remainder of this paper, we focus on the results with an integration time of $t_\mathrm{end}=1000\,P_\p$.

Generally, the maps can be characterized with a stable region at small $a_\m$, an unstable region at larger $a_\m$, and a marginally stable region in between, as can be intuitively expected. There is also a dependence on $i_\pm$, the mutual inclination between the orbit of the moon around the planet, and the orbit of the planet around the inner binary --- typically, high inclinations near $90^\circ$ are less stable than low inclinations (near $0^\circ$ and $180^\circ$), and lead to collisions. The larger instability near high inclinations can be ascribed to Lidov-Kozai (LK) evolution \citep{1962P&SS....9..719L,1962AJ.....67..591K}, and this is explored in more detail below in Sections\,\ref{sect:results:stab:comp} and \ref{sect:results:int}. Also, retrograde orbits ($i_\pm$ near $180^\circ$) tend to be more stable than prograde orbits ($i_\pm$ near $0^\circ$), in the sense that the marginal stability region extends to a larger range in $a_\m$ for retrograde orbits compared to prograde orbits. This is consistent with the general notion that retrograde orbits are typically more stable than prograde orbits owing to the higher relative velocities for retrograde orbits. However, we find that the `binarity' of the stellar binary is also important, especially for retrograde orbits. This is investigated in more detail in \Se\,\ref{sect:results:MMC}.

\subsubsection{Comparison to known results in the single-star case}
\label{sect:results:stab:comp}
In the figures, we show for reference with the vertical blue dashed dotted lines the Hill radius (e.g., \citealt{1992Icar...96...43H}; replacing the binary by a point mass), i.e.,
\begin{align}
\label{eq:r_H}
r_\mathrm{H} = a_\p (1-e_\p) \left(\frac{M_\p}{3(M_1+M_2)} \right )^{1/3}.
\end{align}
Note that $r_\mathrm{H}$ does not lie within the range of $a_\m$ shown in all figures; if the vertical blue dashed dotted line is not visible, then $r_\mathrm{H}$ is larger than the largest value of $a_\m$ shown. Although correct within an order of magnitude, the Hill radius only captures the true stability boundary within a factor of a few. This is not very surprising given that the Hill radius strictly only applies to the three-body case (with the binary replaced by a single body). 

In the limit that the binary can be interpreted as a point mass, our results can be compared to \citet{2017MNRAS.466..276G}, who considered the stability of inclined hierarchical three-body systems. \citet{2017MNRAS.466..276G} find the following expression for the limiting radii of stability as a function of the mutual inclination (in our case, $i_\pm$),
\begin{align}
\label{eq:grishin}
r_\mathrm{c}^{\mathrm{LK}}(i_\pm) = f_\mathrm{fudge} \, r_\mathrm{H} \, g(i_\pm)^{-2/3} p(i_\pm).
\end{align}
Here,
\begin{align}
g(i_\pm) = \cos(i_\pm) + \sqrt{ 3 + \cos^2 (i_\pm)},
\end{align}
and
\begin{align}
\nonumber p(i_\pm) &= 0.35161117i_\pm^4 - 2.431451i_\pm^3  +  6.54177136i_\pm^2 \\
&\quad - 8.01396441i_\pm +  4.40019183
\end{align}
for $0.867<i_\pm < 2.41$; otherwise, $p(i_\pm)=1$\footnote{Our adopted polynomial fit is different from Table 1 of \citealt{2017MNRAS.466..276G}; there is an error in the latter table (E. Grishin, private communication).}. In defining the function $p(i_\pm)$, $i_\pm$ is measured in radians. The `fudge factor' $f_\mathrm{fudge}$ is set to $f_\mathrm{fudge} = 0.85$, in order to match with our results for Kepler 16 at $i_\pm=0^\circ$. We remark that the Hill radius is not a true measure of stability but a proxy, that there are no unique criteria for `stability', and that there is uncertainty in the fudge factor. 

\begin{figure*}
\center
\includegraphics[scale = 0.395, trim = 35mm 3mm 0mm 10mm]{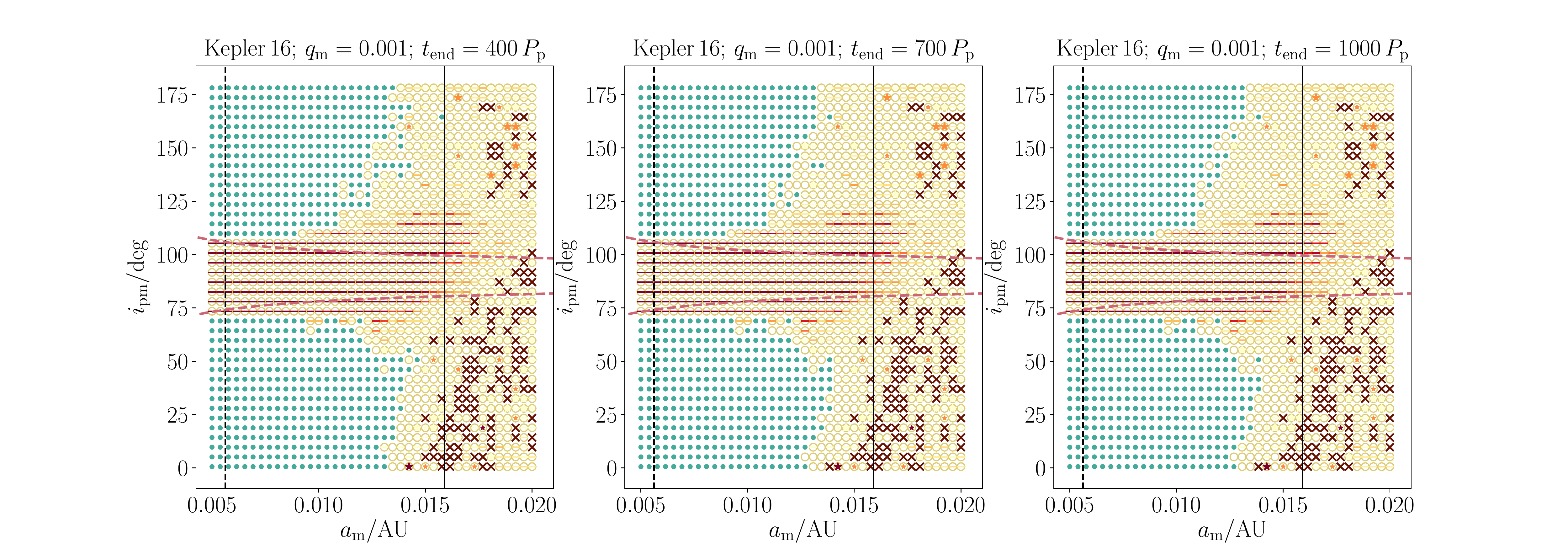}
\includegraphics[scale = 0.395, trim = 35mm 3mm 0mm 0mm]{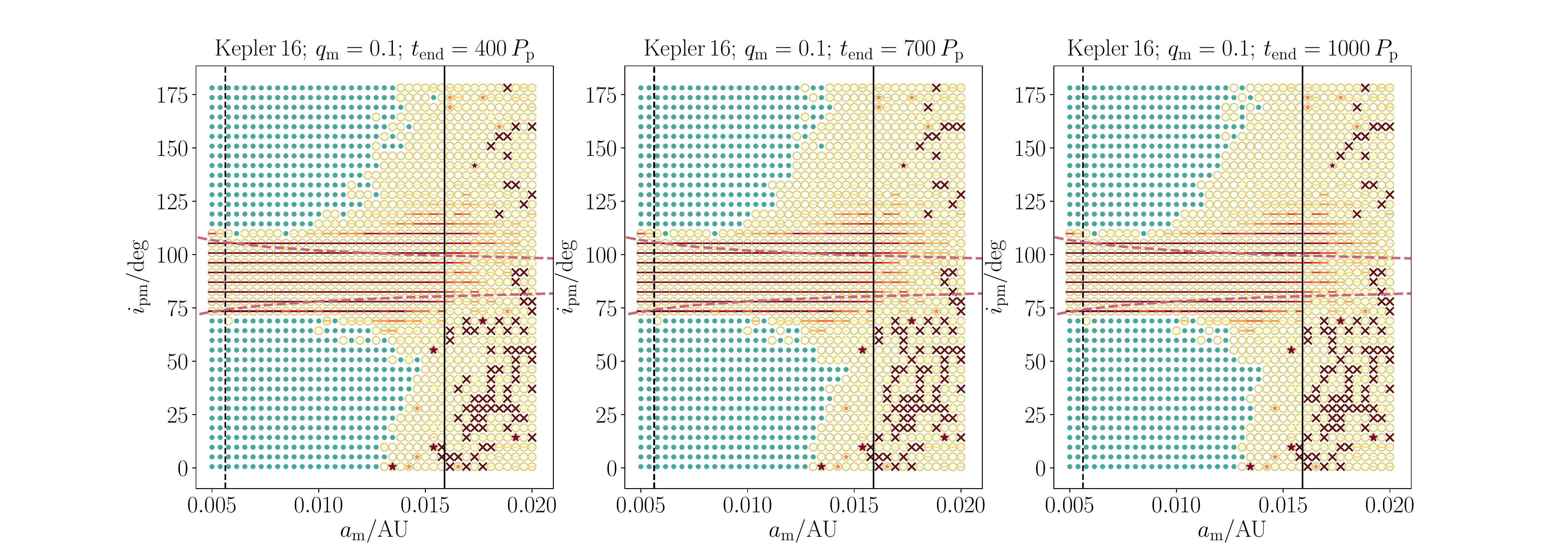}
\includegraphics[scale = 0.395, trim = 35mm 10mm 0mm 0mm]{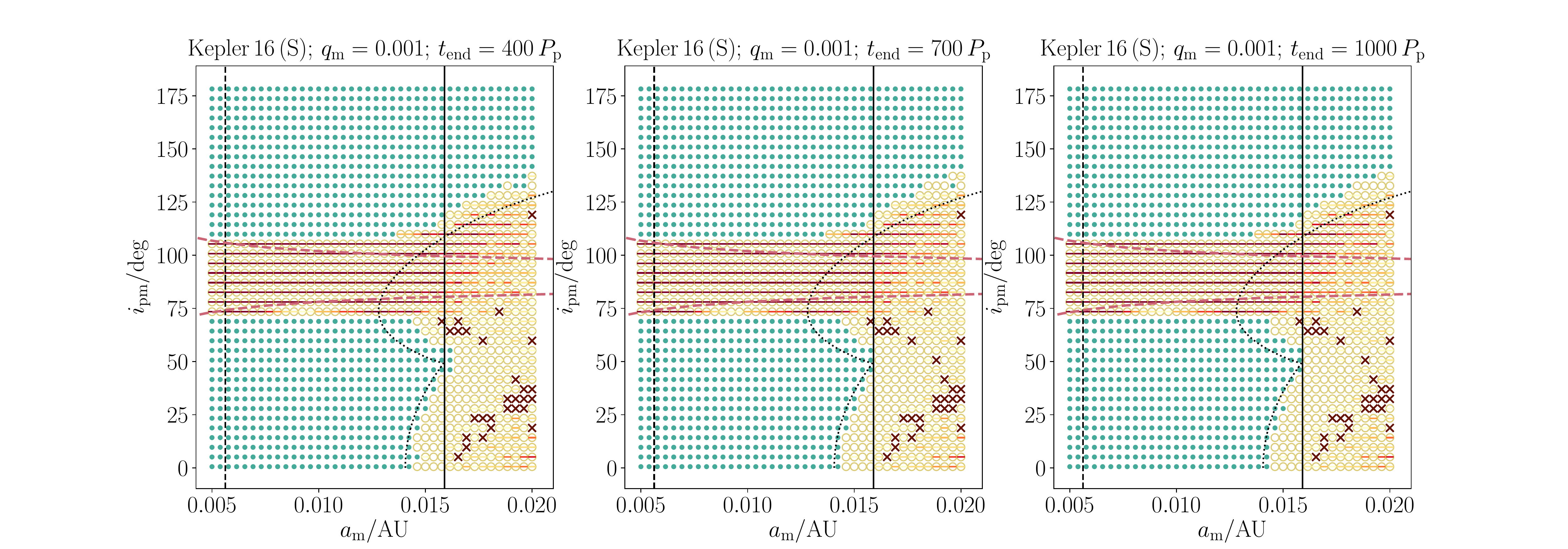}
\caption{\small Stability maps for the Kepler 16 system. In each panel, each point represents 10 realizations with different initial lunar mean anomaly, which are used to define stable (green filled circles), completely unstable (red crosses), and marginally stable (yellow open circles) systems. Collisions of the moon with the planet and stars are indicated with `$-$' and `$\star$', respectively, where the color encodes the frequency with respect to the 10 realizations of the initial lunar mean anomaly (yellow to red for 1 to 10). Here, we show the frequency for the most common collision type among the mean anomalies. For collisions of the moon with the stars, the large and small `$\star$' symbols correspond to collisions with the primary and secondary star, respectively. The three panels in each row correspond to different integration times: 400, 700 and 1000 $P_\p$. In the first and third columns, the mass ratio of the moon to the planet is $q_\m = M_\m/M_\p = 0.001$, whereas in the second column, $q_\m = 0.1$. The third row corresponds to the single-star case (`S'), in which the stellar binary is replaced by a point mass. The vertical black solid lines show the location of the 1:1 MMC of the moon with the binary (equation~\ref{eq:a_MMC}), and the vertical black dashed lines show the locations of the 2:1 MMC. The red dashed lines show the boundary for collisions of the moon with the planet due to LK evolution, computed according to equation~(\ref{eq:e_max_LK}). In the third row, the black dotted lines show equation~(\ref{eq:grishin}), which is a polynomial for the single-star case adopted from \citet{2017MNRAS.466..276G}. }
\label{fig:stab_k16}
\end{figure*}

\begin{figure*}
\center
\includegraphics[scale = 0.34, trim = 0mm -10mm 0mm 12mm]{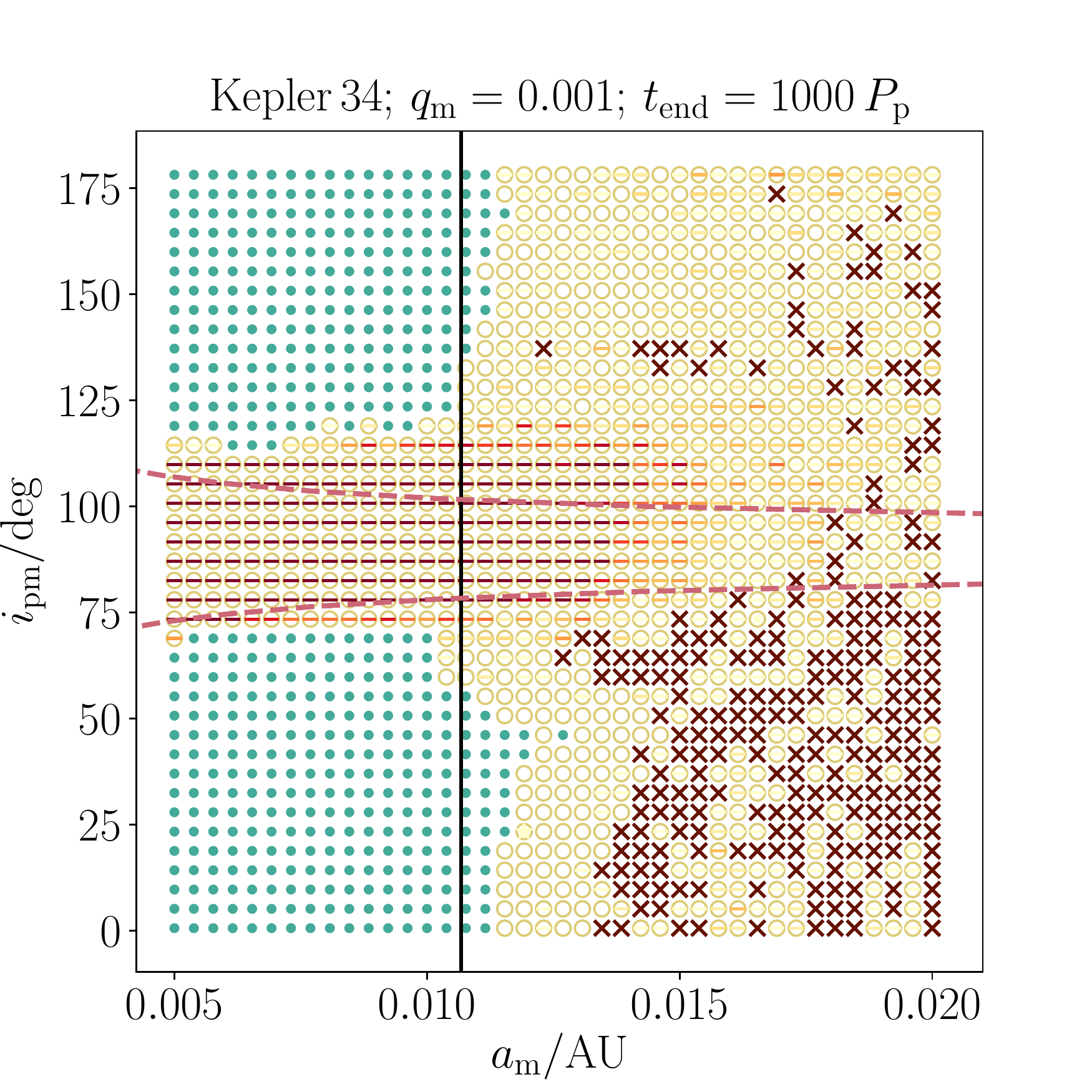}
\includegraphics[scale = 0.34, trim = 15mm -10mm 0mm 12mm]{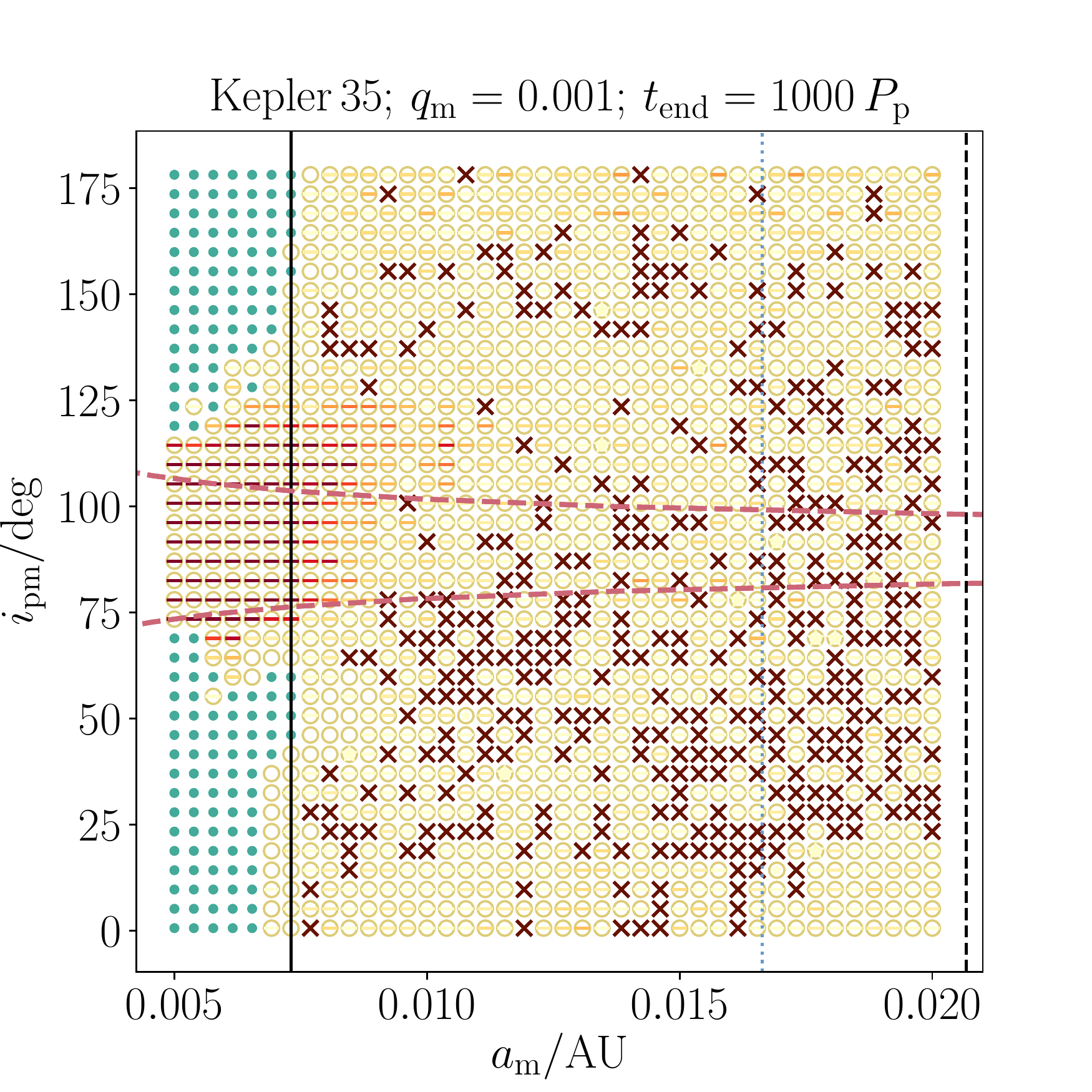}
\includegraphics[scale = 0.34, trim = 15mm -10mm 0mm 12mm]{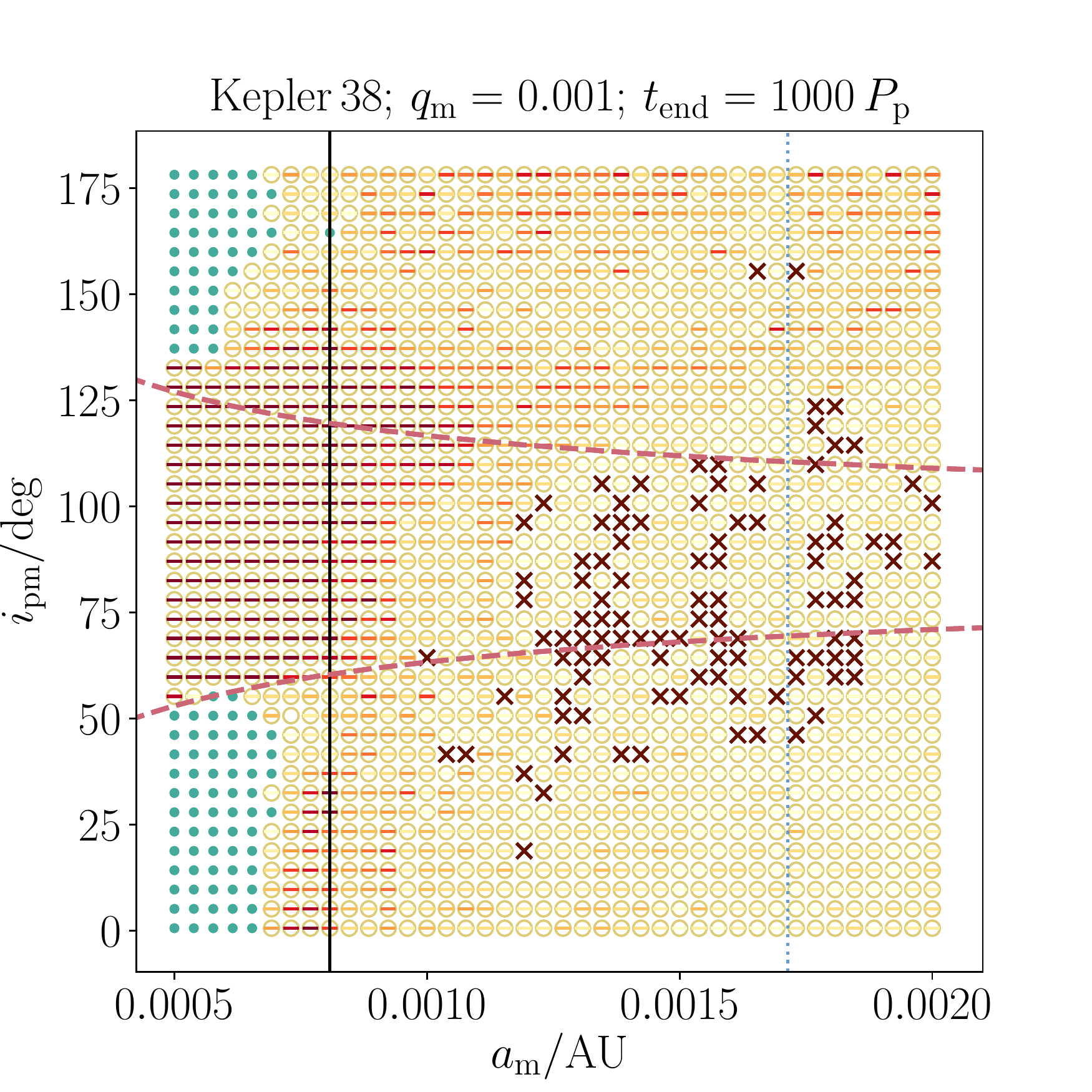}
\includegraphics[scale = 0.34, trim = 0mm -10mm 0mm 12mm]{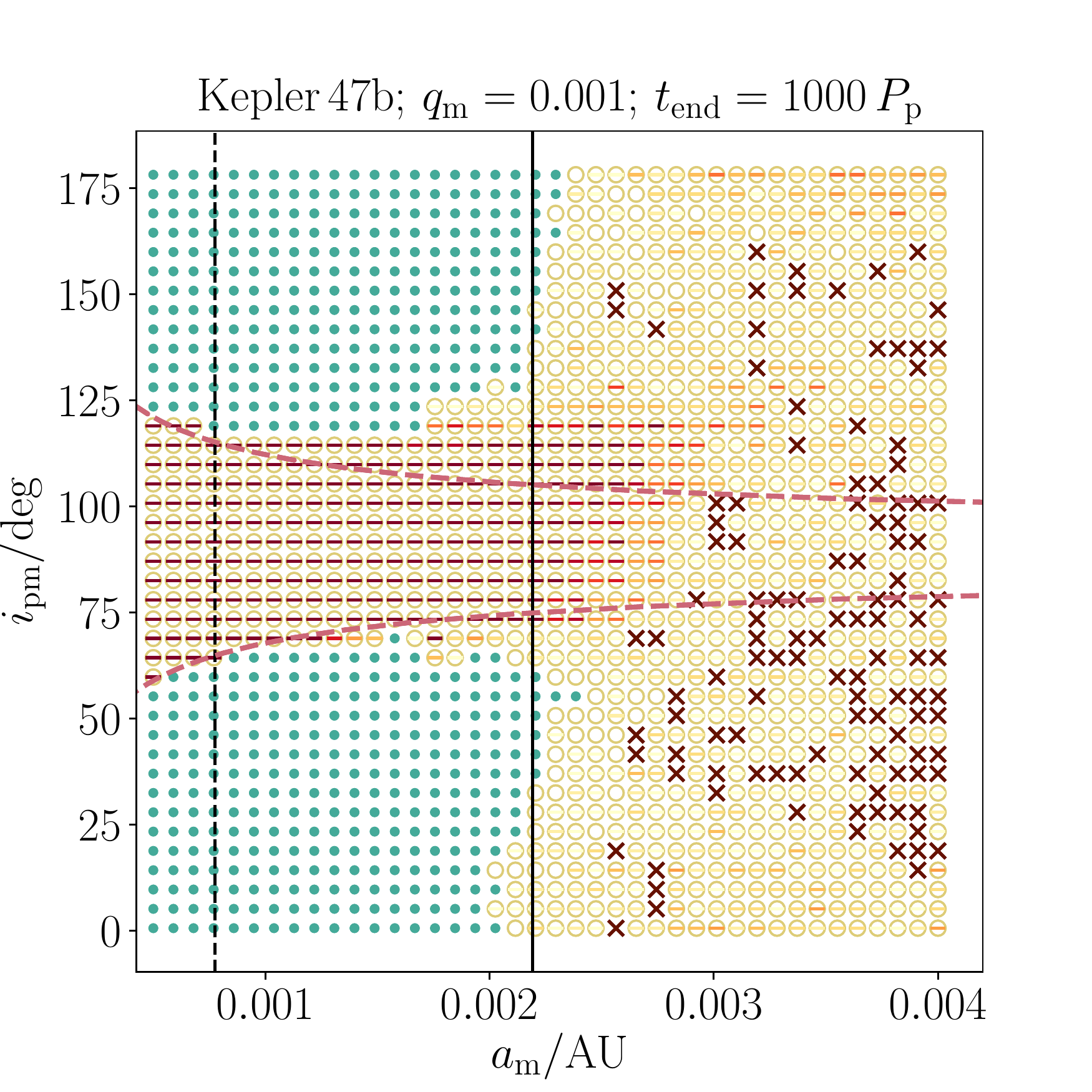}
\includegraphics[scale = 0.34, trim = 15mm -10mm 0mm 12mm]{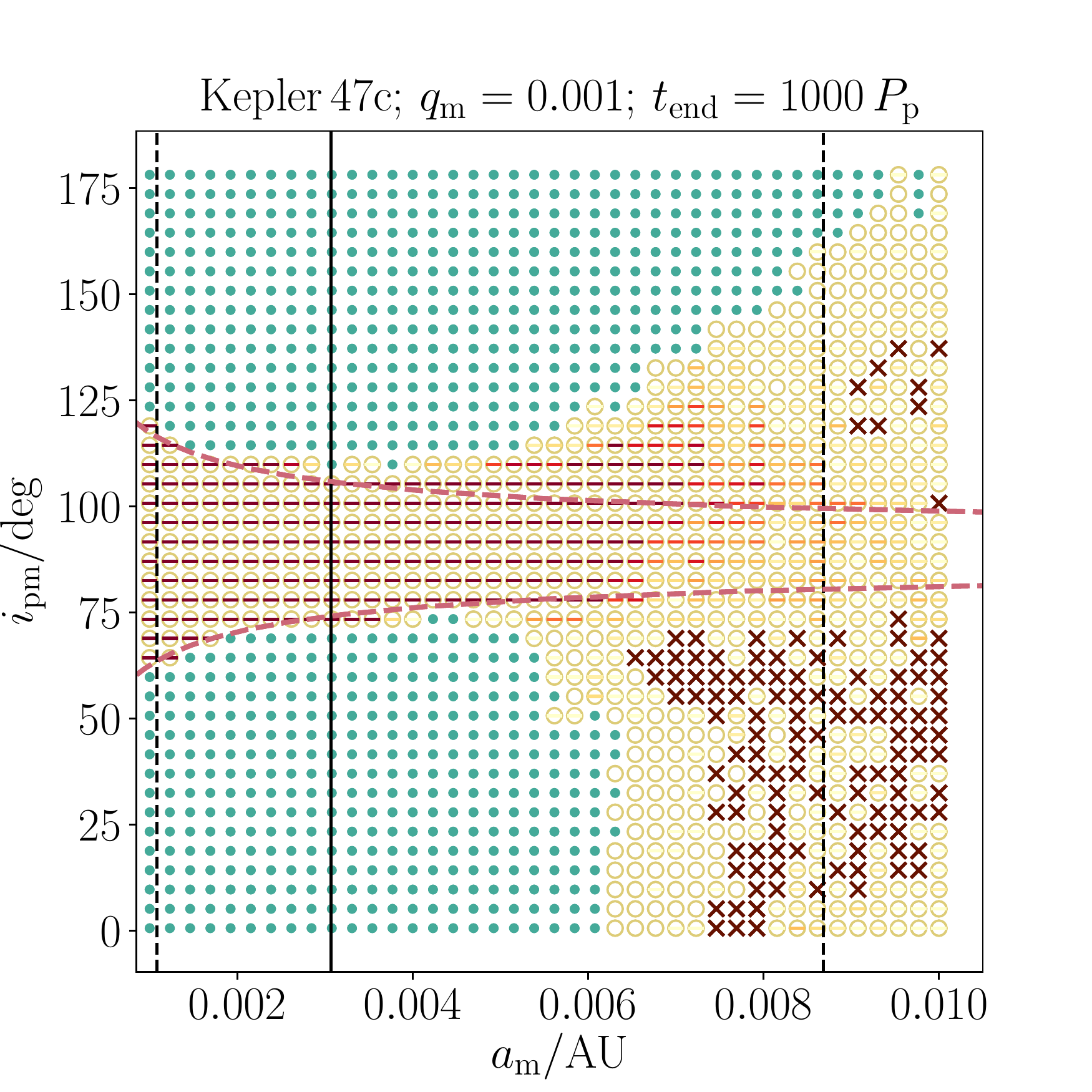}
\includegraphics[scale = 0.34, trim = 15mm -10mm 0mm 12mm]{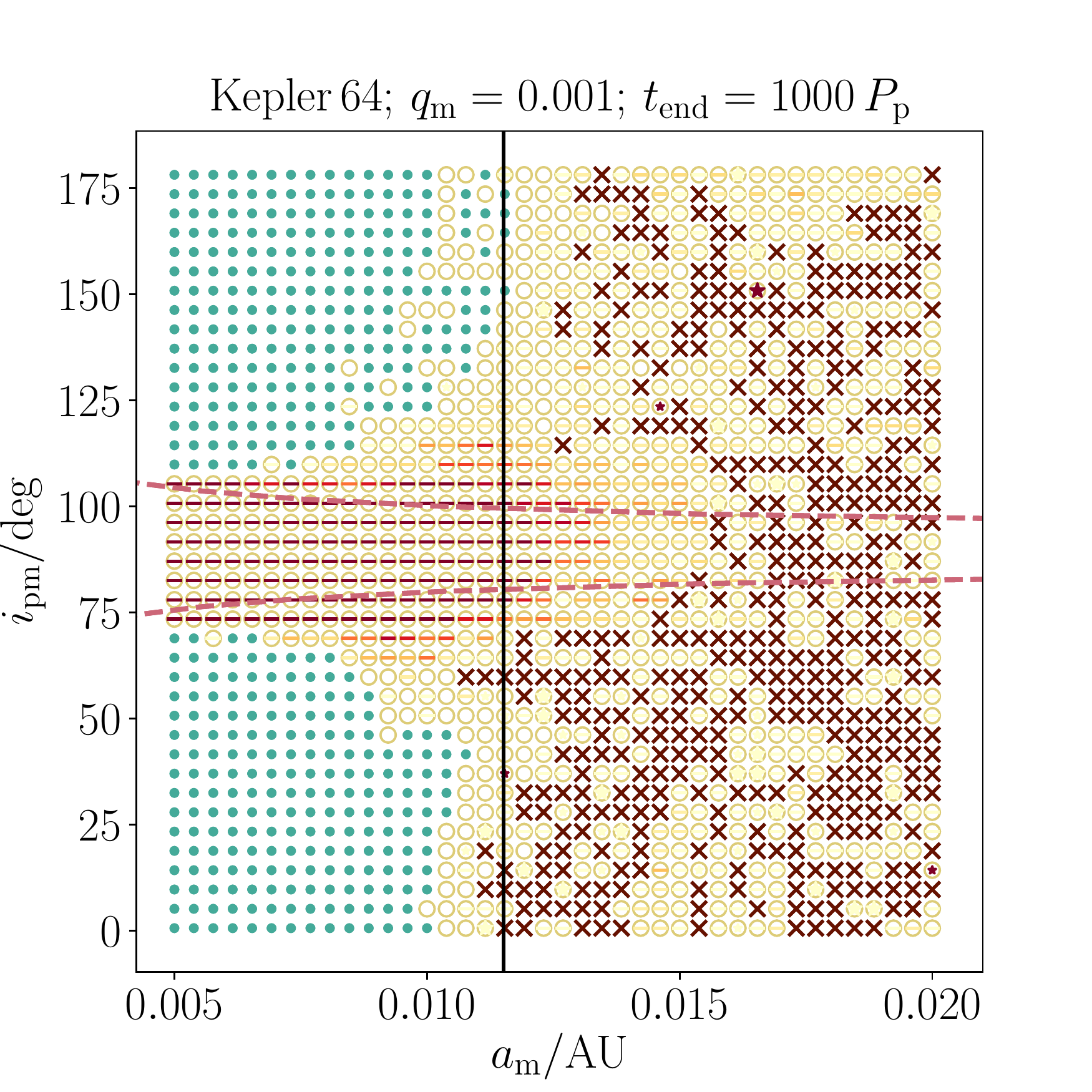}
\includegraphics[scale = 0.34, trim = 0mm -10mm 0mm 12mm]{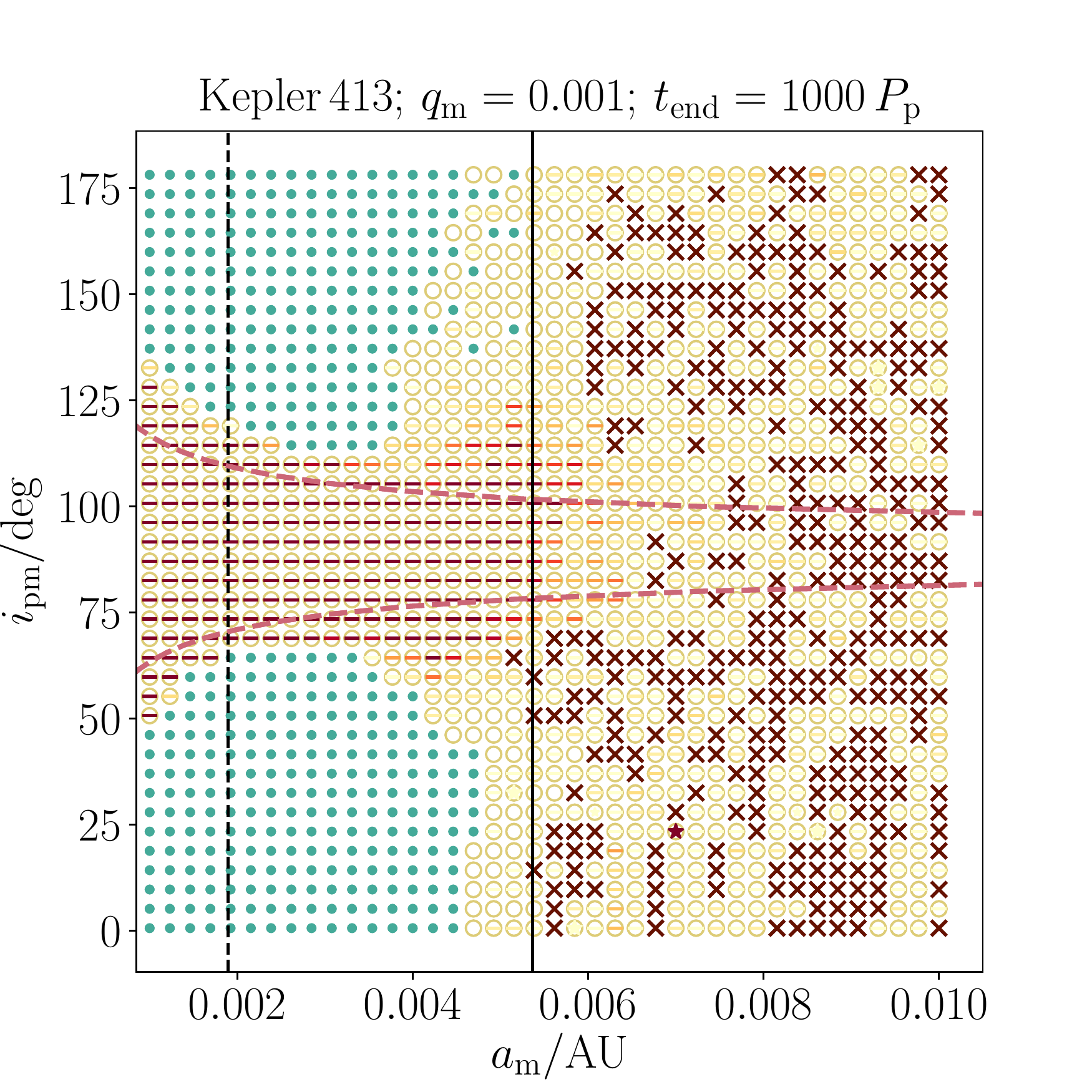}
\includegraphics[scale = 0.34, trim = 15mm -10mm 0mm 12mm]{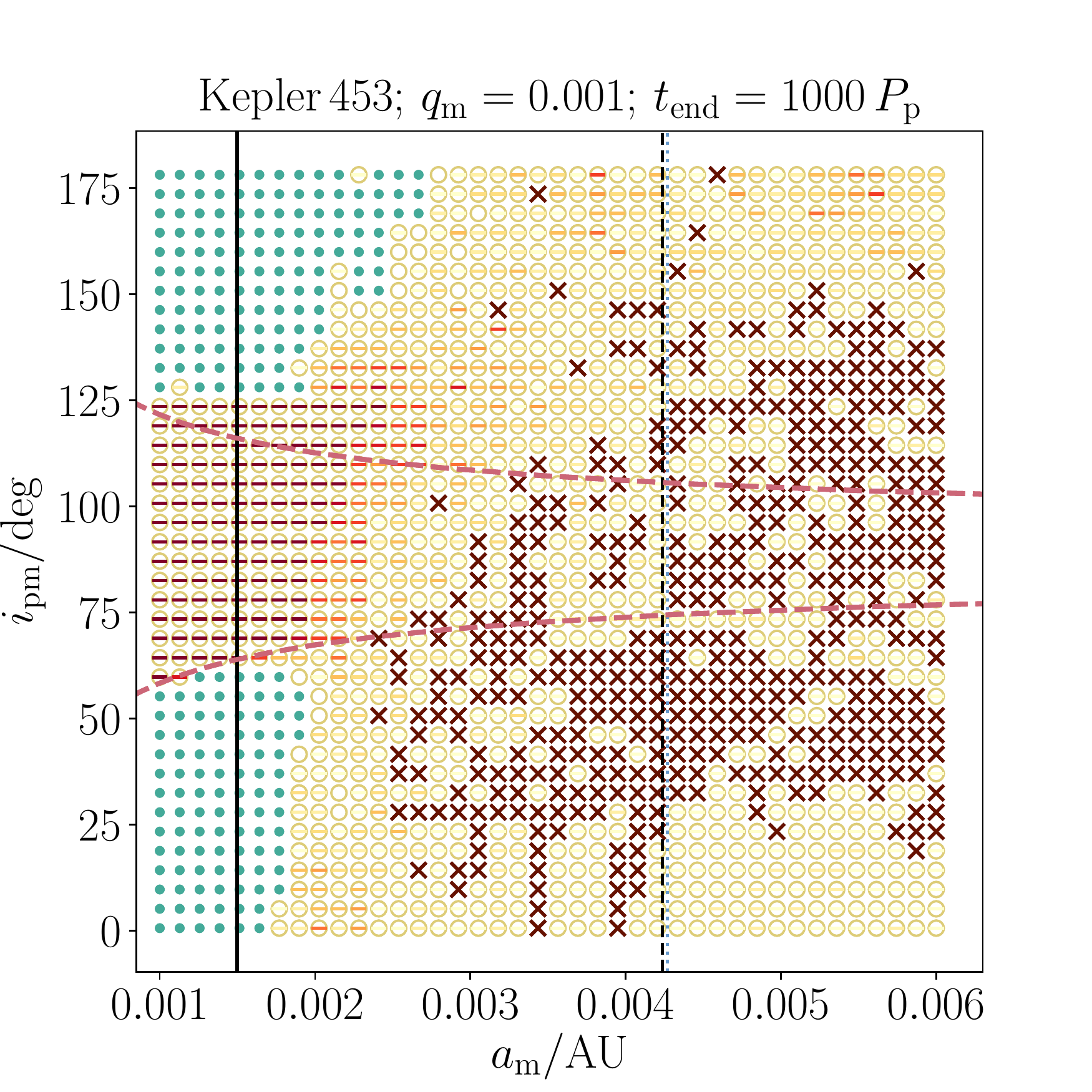}
\includegraphics[scale = 0.34, trim = 15mm -10mm 0mm 12mm]{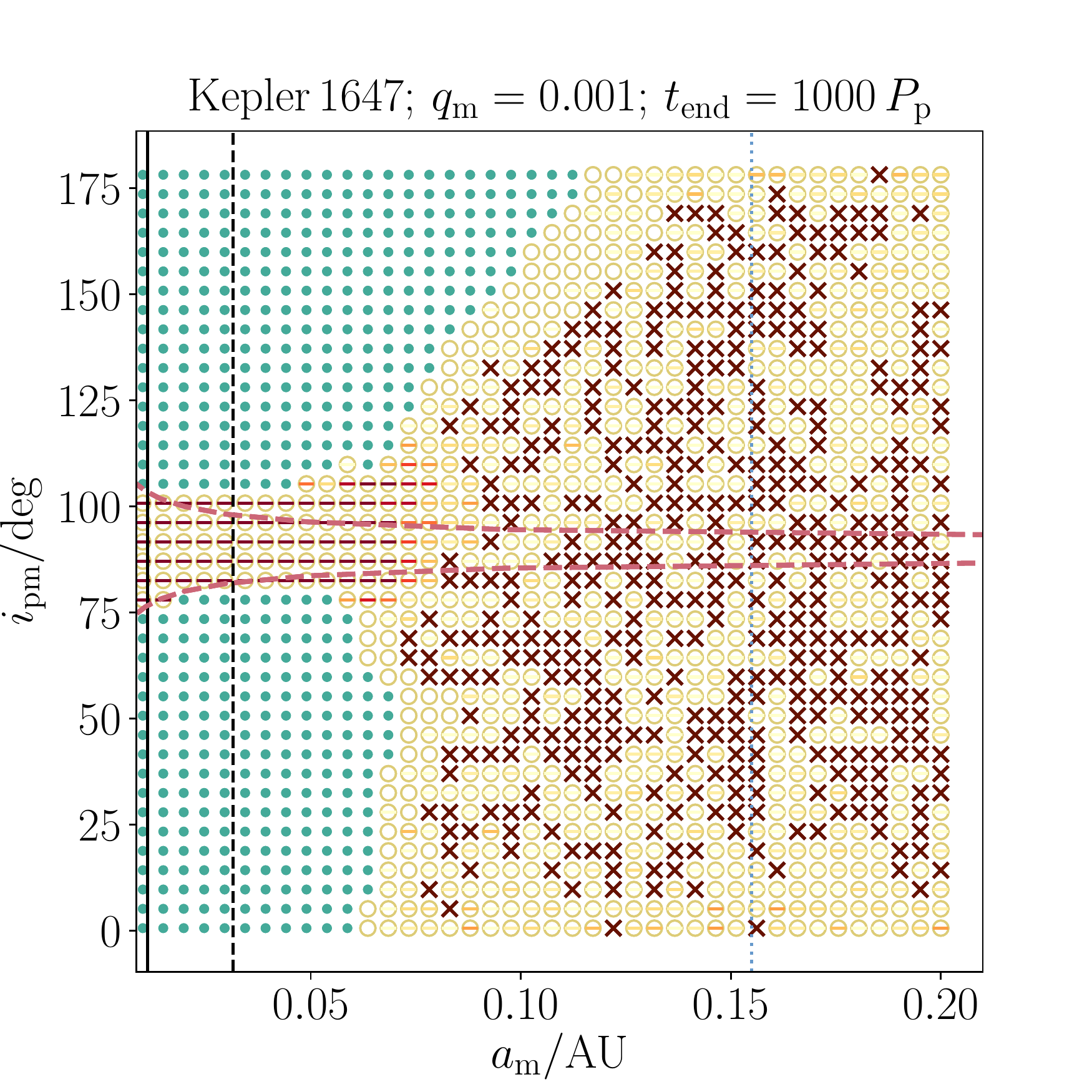}
\caption{\small Stability maps for nine {\it Kepler} CBP systems, similar to \F\,\ref{fig:stab_k16}. The name of the system is indicated above each panel. In all cases, $q_\m = 0.001$, and the integration time is 1000 $P_\p$. The vertical black solid lines show the location of the 1:1 MMC of the moon with the binary (equation~\ref{eq:a_MMC}); the vertical black dashed lines show the locations of the 2:1 and 1:2 MMCs. The red dashed lines show the boundary for collisions of the moon with the planet due to LK evolution, computed according to equation~(\ref{eq:e_max_LK}). The vertical blue dashed lines (not visible in all panels) show the Hill radius, equation~(\ref{eq:r_H}).}
\label{fig:stab_krest}
\end{figure*}

\begin{figure*}
\center
\includegraphics[scale = 0.34, trim = 0mm -10mm 0mm 12mm]{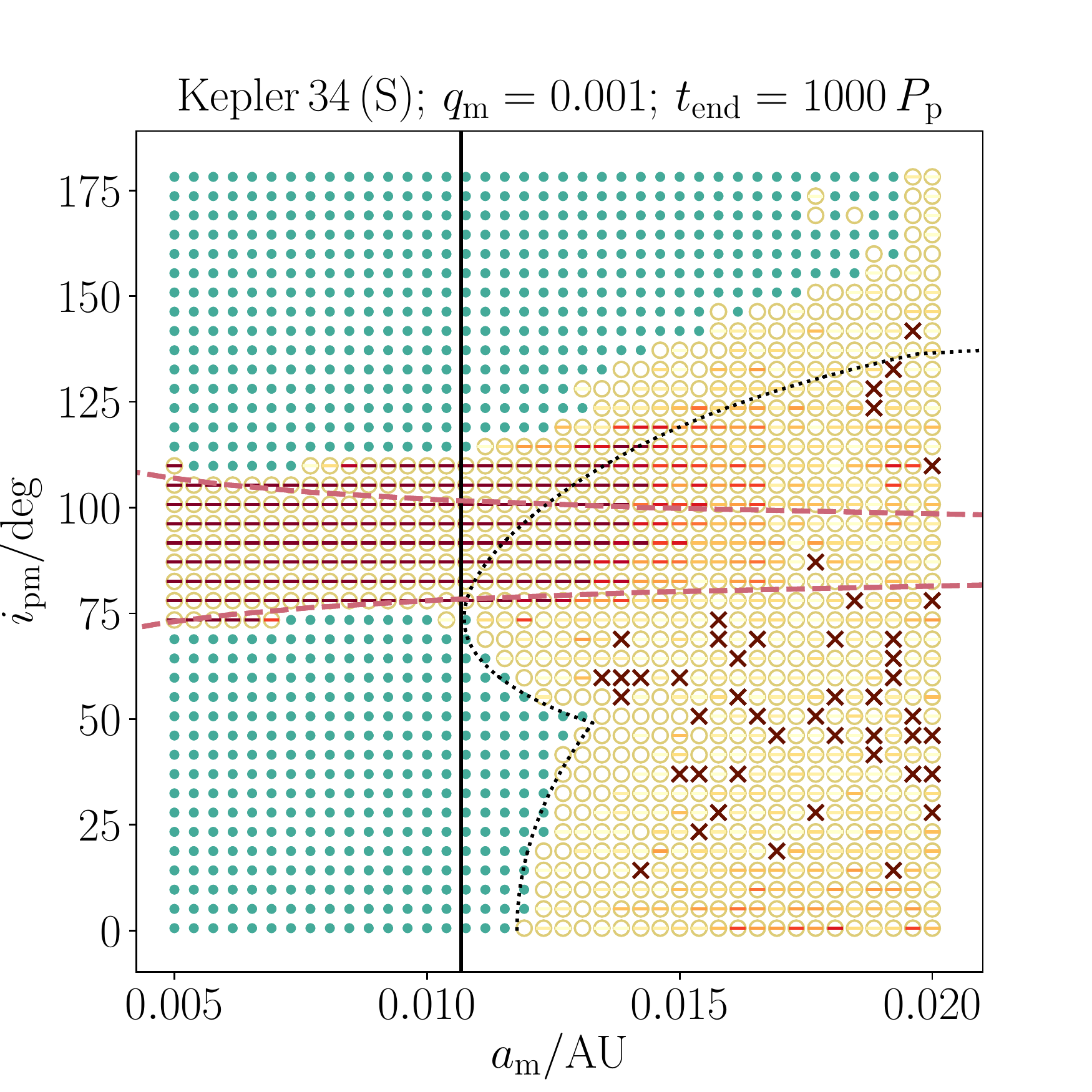}
\includegraphics[scale = 0.34, trim = 15mm -10mm 0mm 12mm]{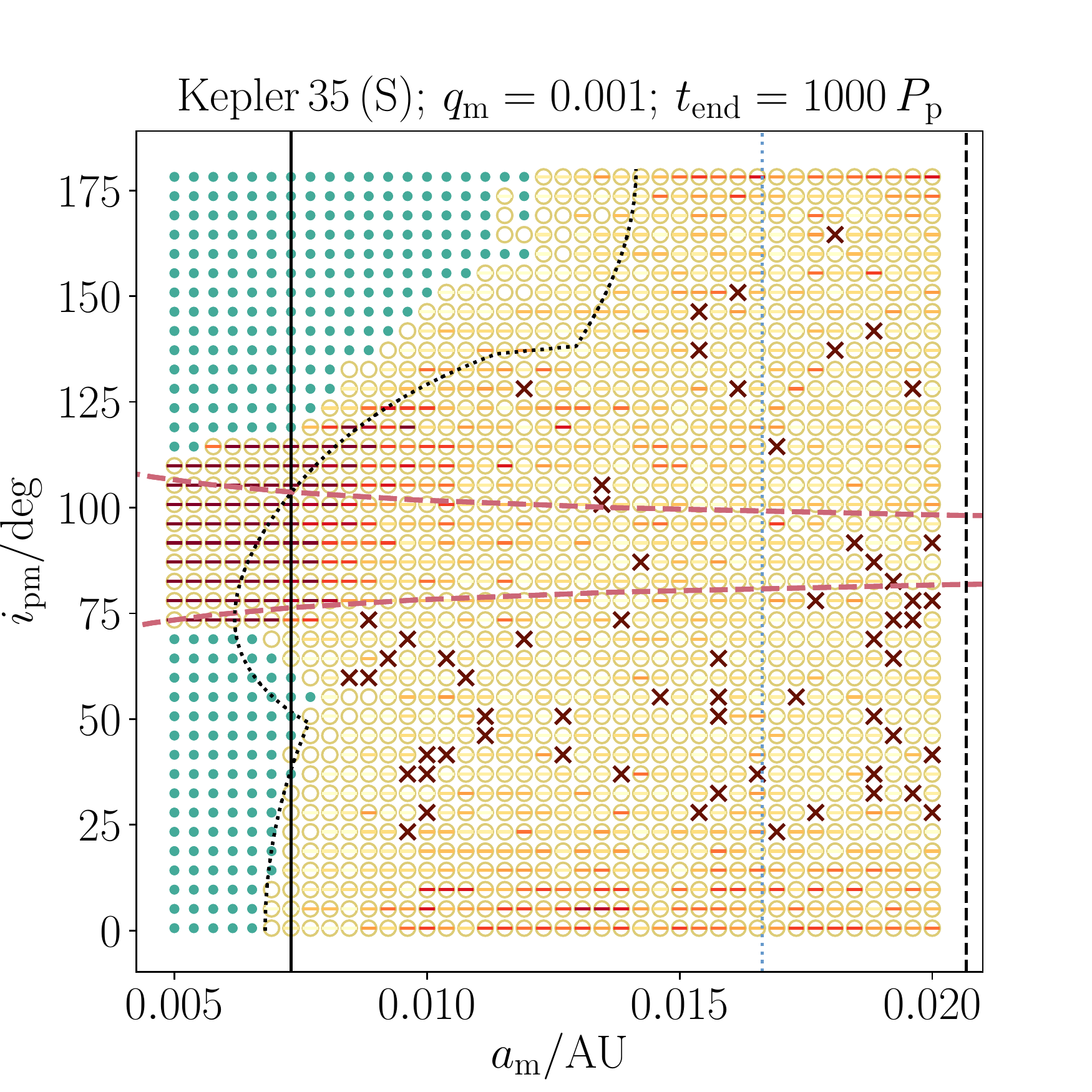}
\includegraphics[scale = 0.34, trim = 15mm -10mm 0mm 12mm]{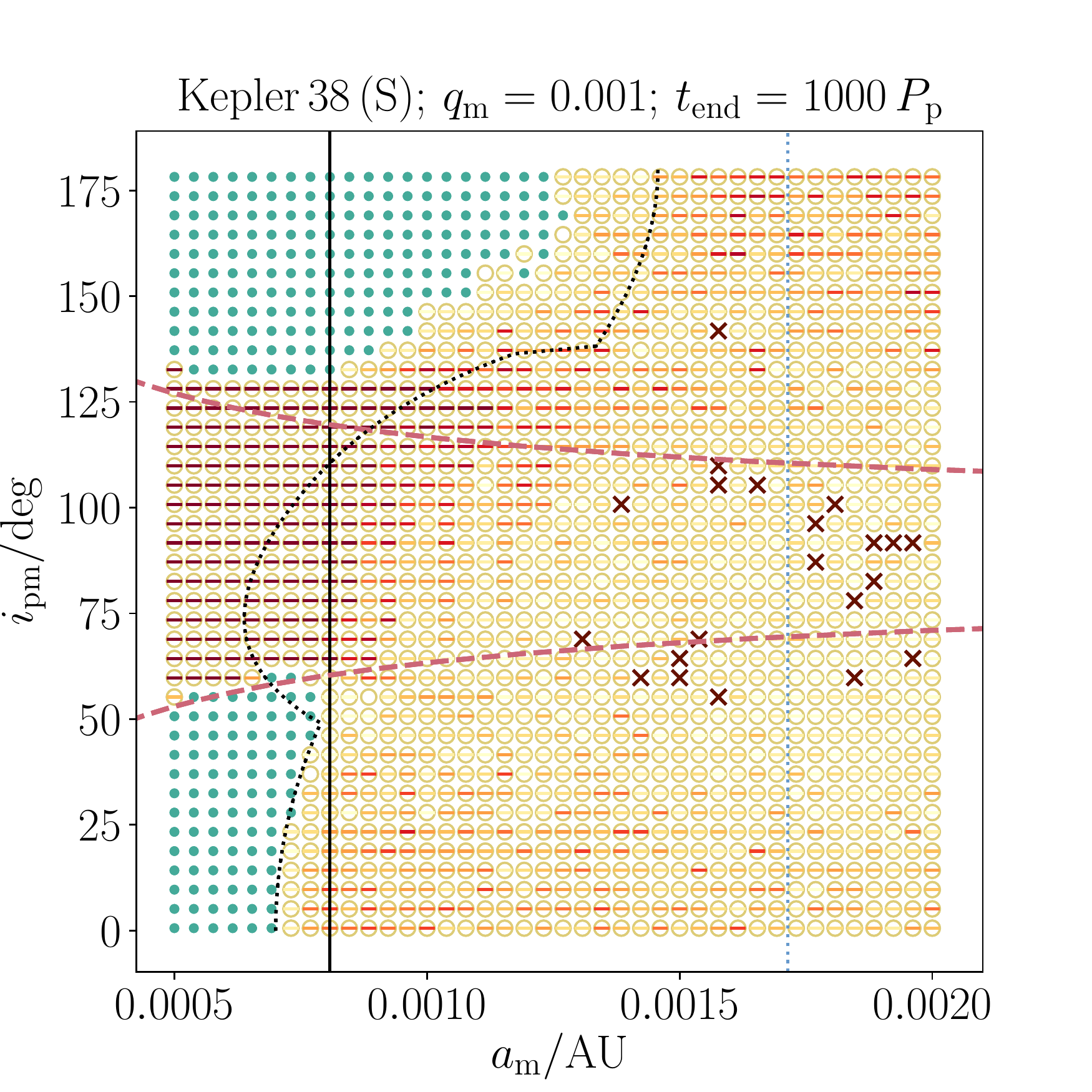}
\includegraphics[scale = 0.34, trim = 0mm -10mm 0mm 12mm]{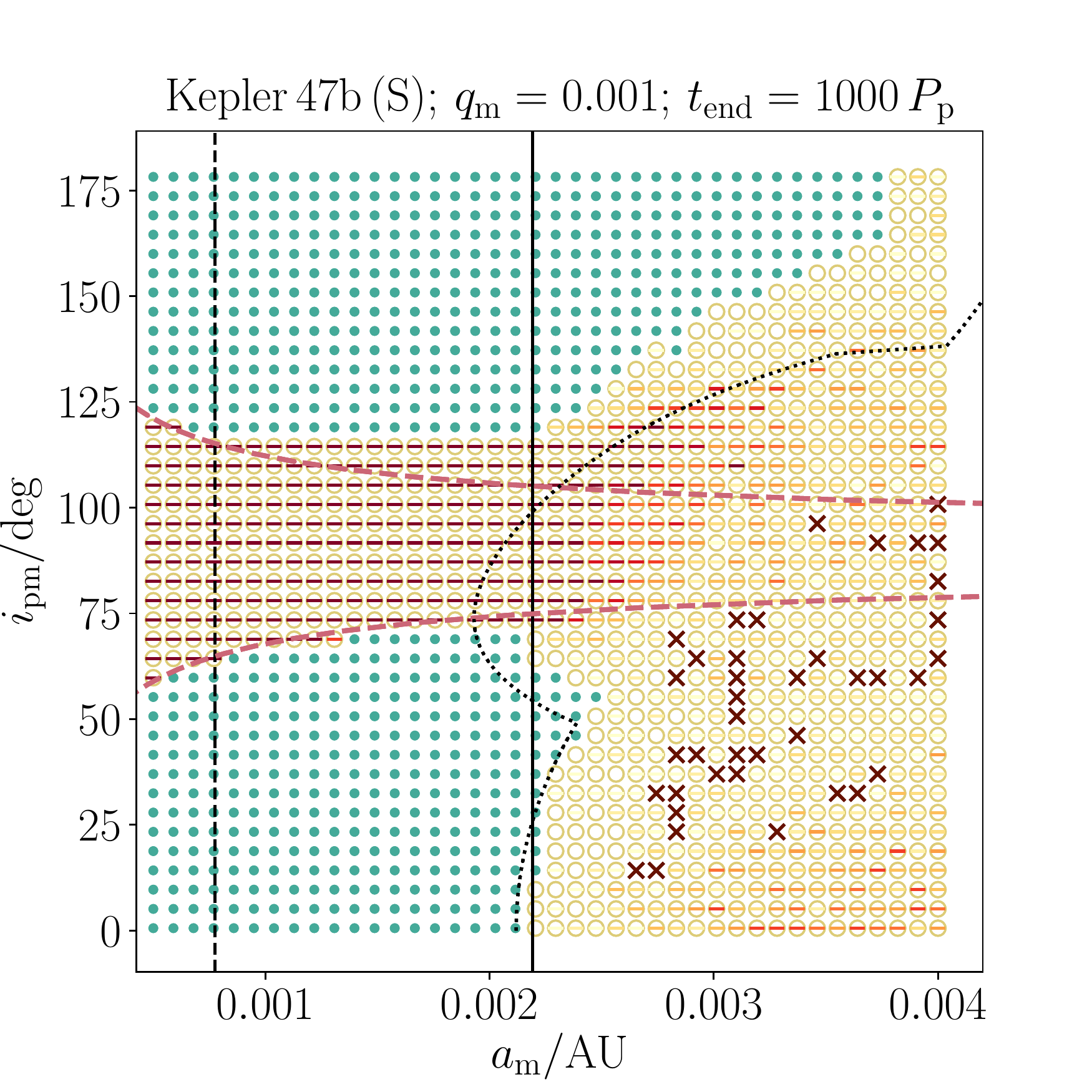}
\includegraphics[scale = 0.34, trim = 15mm -10mm 0mm 12mm]{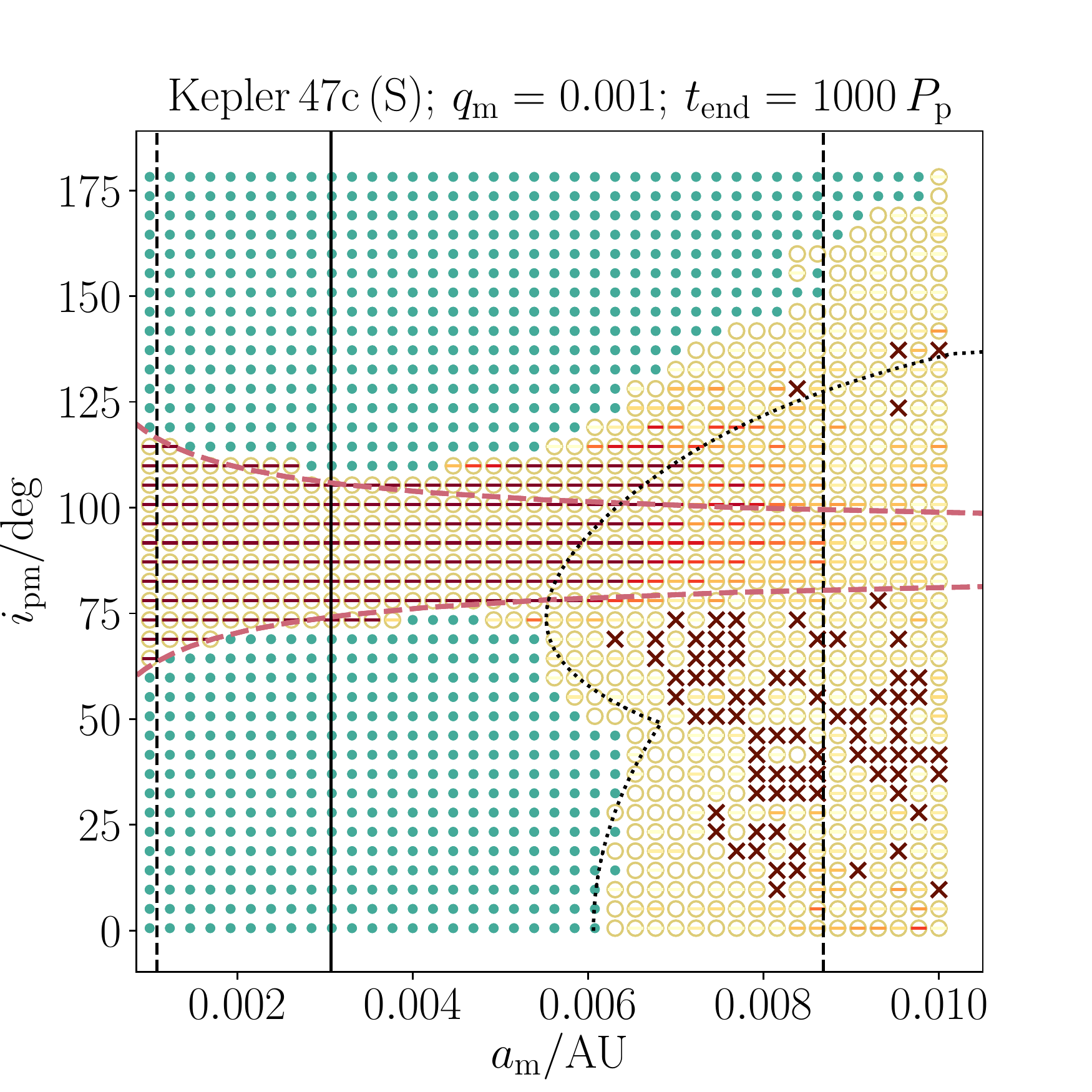}
\includegraphics[scale = 0.34, trim = 15mm -10mm 0mm 12mm]{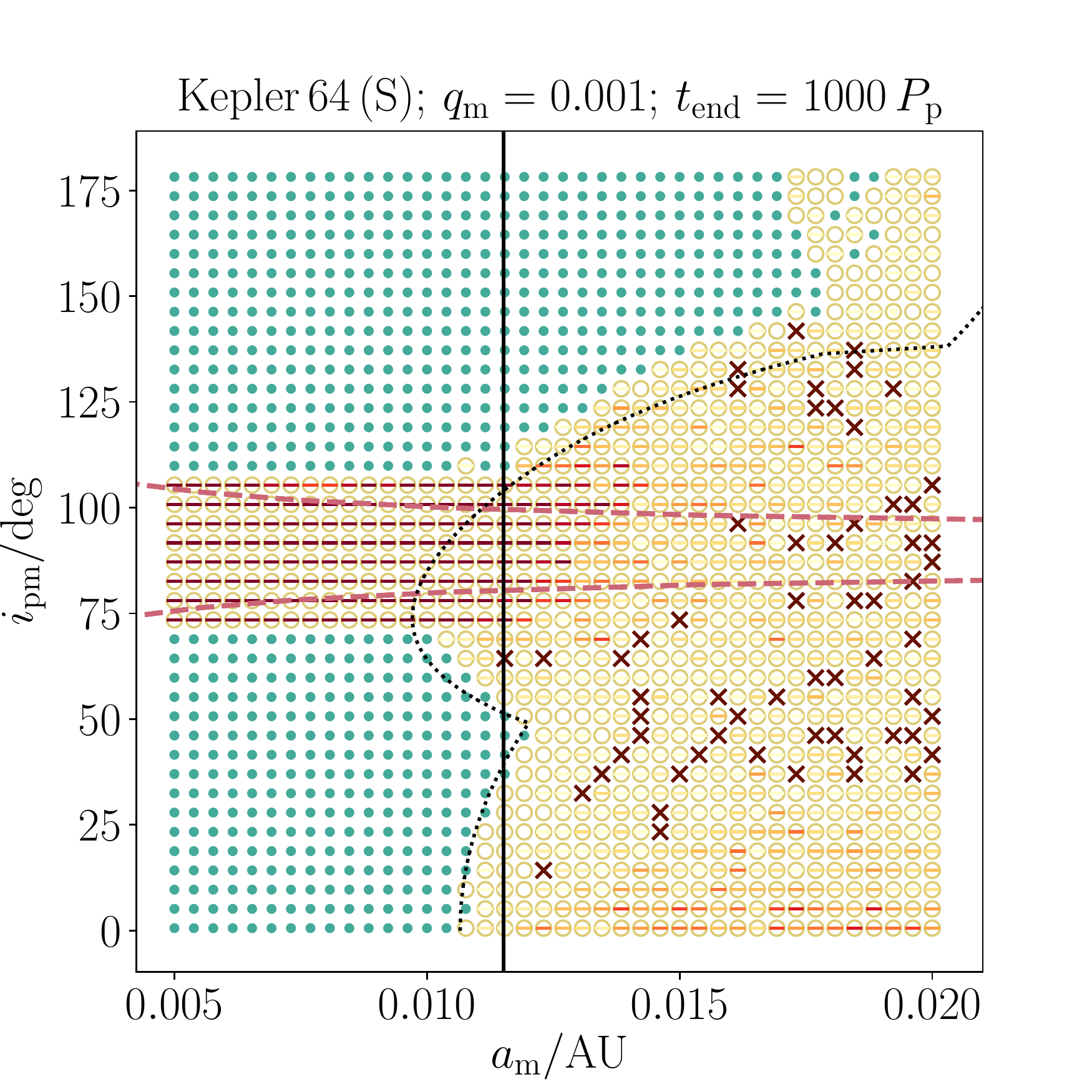}
\includegraphics[scale = 0.34, trim = 0mm -10mm 0mm 12mm]{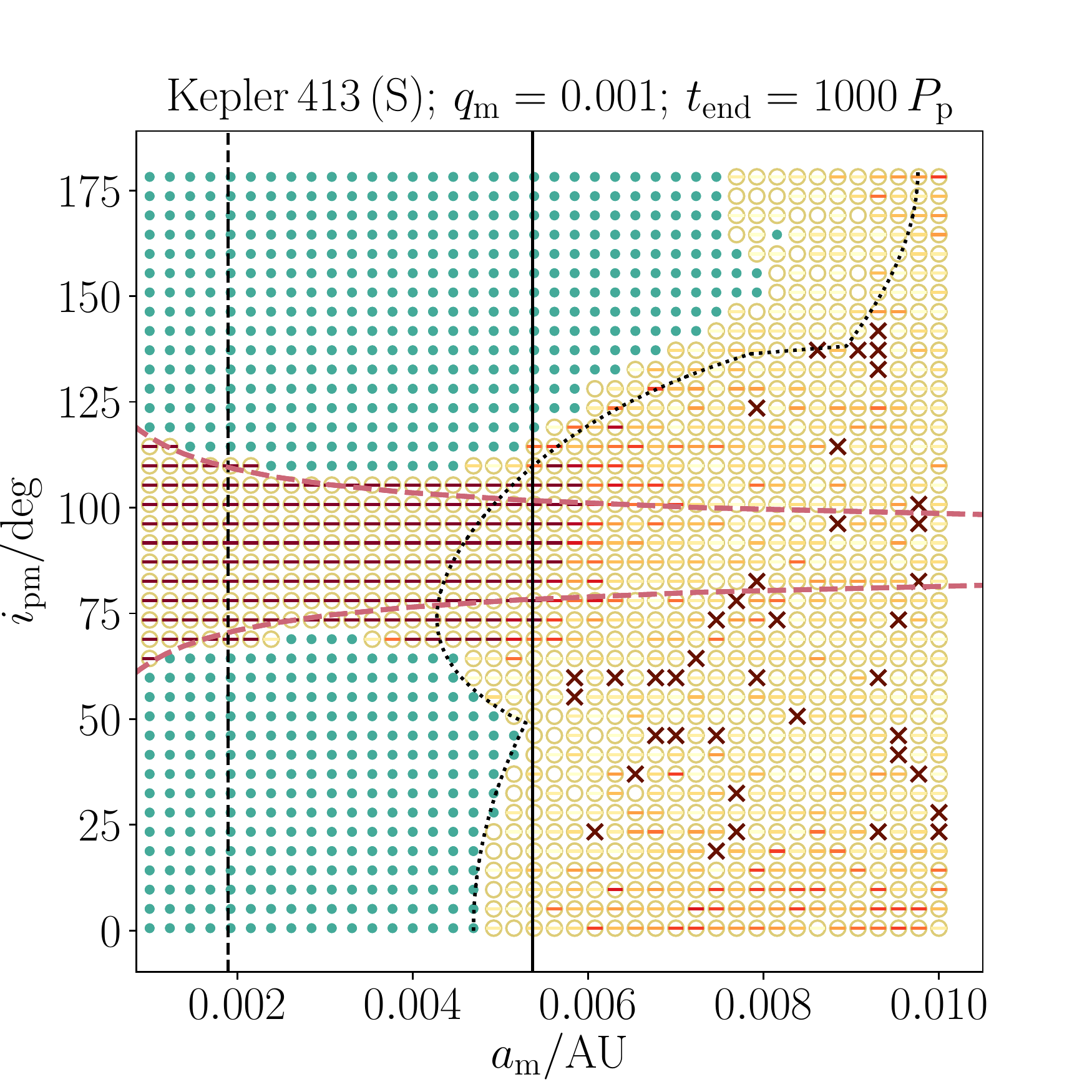}
\includegraphics[scale = 0.34, trim = 15mm -10mm 0mm 12mm]{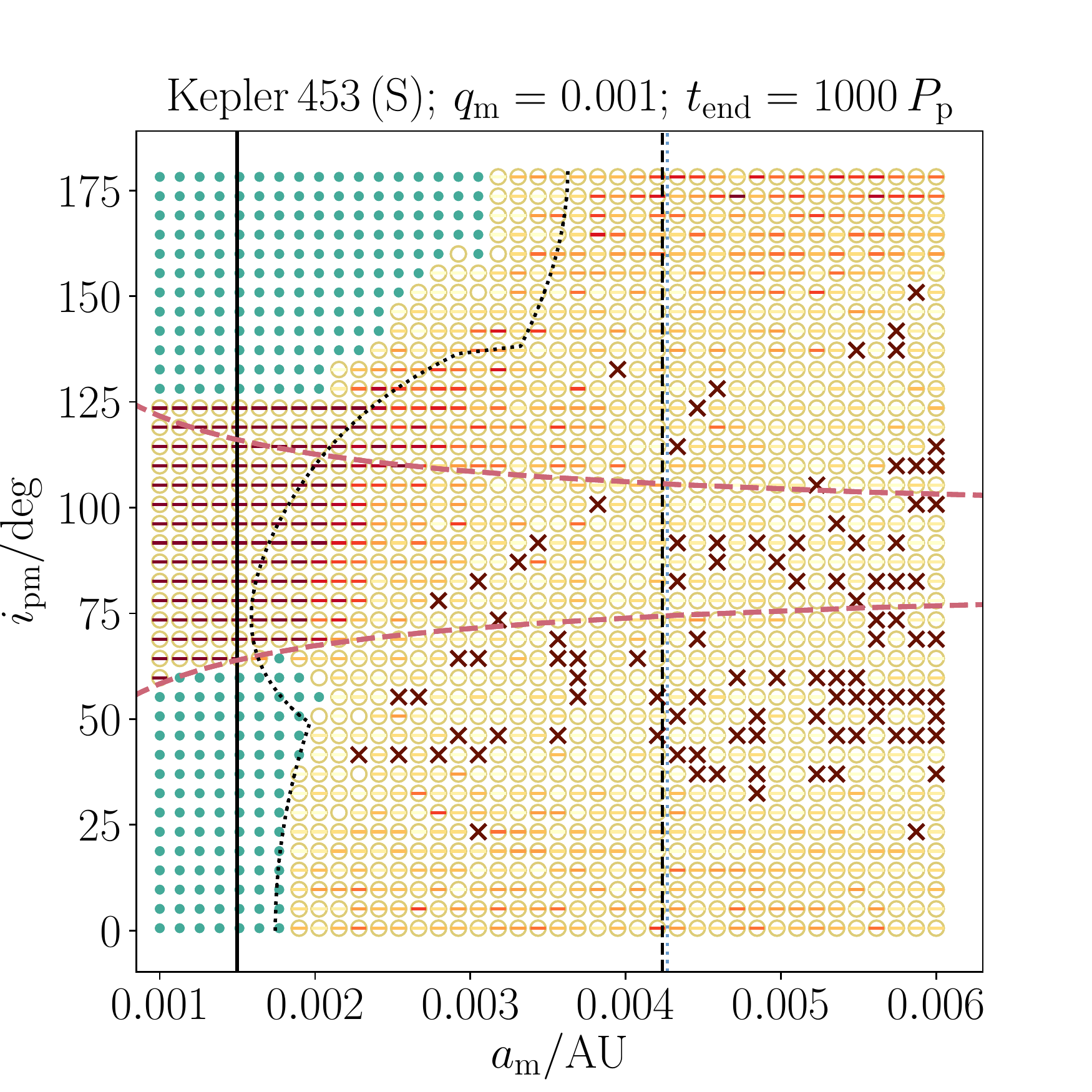}
\includegraphics[scale = 0.34, trim = 15mm -10mm 0mm 12mm]{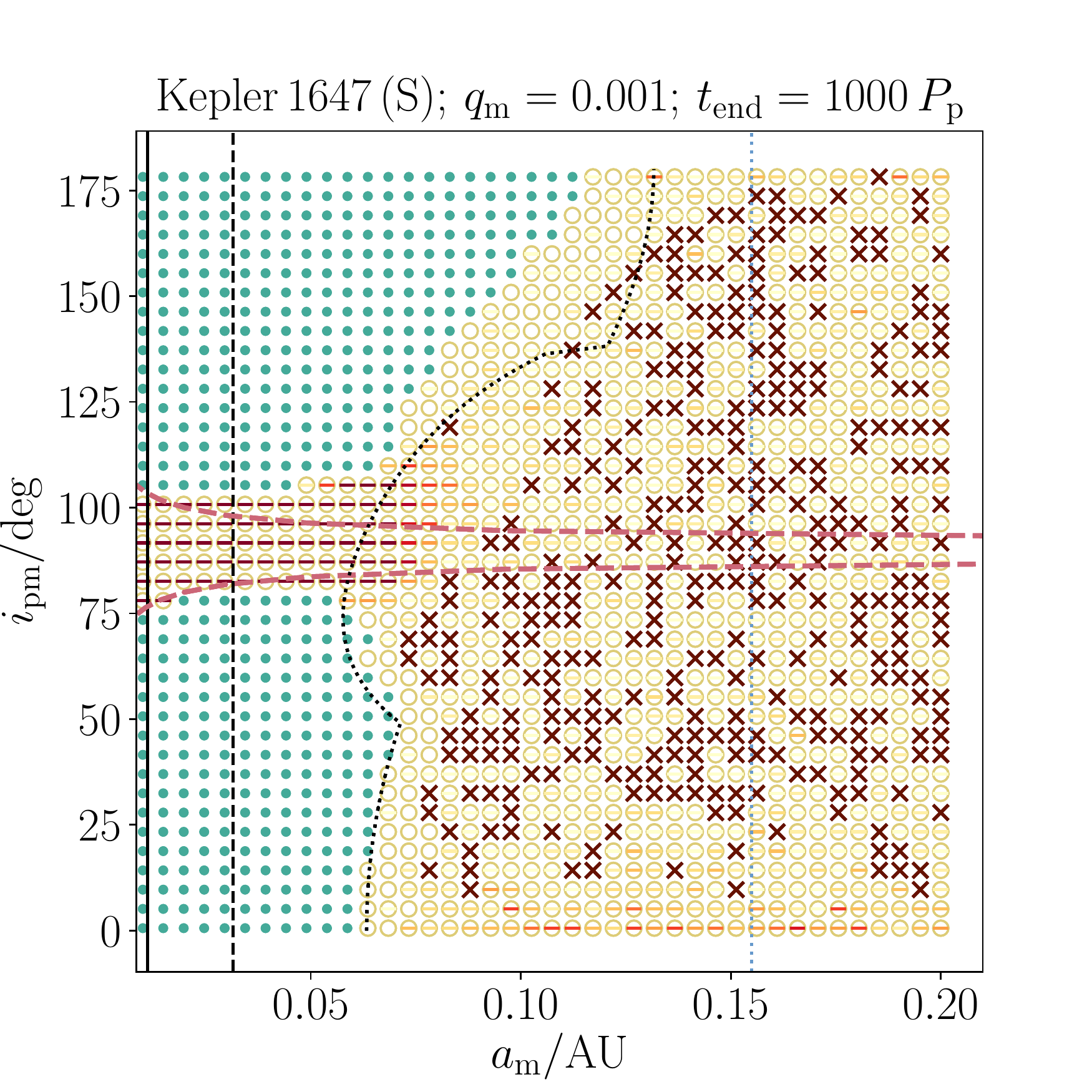}
\caption{\small Similar to \F\,\ref{fig:stab_krest}, now for the single-star case (`S') and $q_\m = 0.001$. The black dotted lines show equation~(\ref{eq:grishin}), which is a polynomial for the single-star case adopted from \citet{2017MNRAS.466..276G}. }
\label{fig:stab_krest_single}
\end{figure*}

For the single-star cases, we show equation~(\ref{eq:grishin}) with the black dotted lines in the third row of \F\,\ref{fig:stab_k16} and in \F\,\ref{fig:stab_krest_single} for Kepler 16 and the other {\it Kepler} CBP systems, respectively. For all systems, the absolute locations of the stability boundaries in the simulations agree well at low inclinations. At higher inclinations, in particular near retrograde orientations, the fits tend to give larger boundaries in terms of $a_\m$ compared to the simulations. This can be attributed to the non-Keplerian but chaotic nature of orbits beyond $\sim 0.6\,r_\mathrm{H}$, implying that the averaging method fails \citep{2017MNRAS.466..276G}. In addition, \citet{2017MNRAS.466..276G} use a different definition for the stability of orbits. Qualitatively, the general shapes of the boundaries agree, with the $a_\m$-boundaries increasing with increasing $i_\pm$ (i.e., retrograde orbits are more stable in the single-star case), and a `bulge' near high inclination. Note that equation~(\ref{eq:grishin}) does not take into account collisions of the moon with the planet, which determine the stability boundaries in the simulations for inclinations near $90^\circ$. The latter regime is discussed in more detail in \Se\,\ref{sect:results:int}.

\subsection{Mean motion commensurabilities}
\label{sect:results:MMC}
As mentioned above, the Hill radius does not accurately describe the boundary between stable and unstable orbits in the $(a_\m,i_\pm)$ plane. Here, we consider an alternative simple analytic description of the stability boundary using an interpretation based on commensurability of the mean motions of the binary and the moon. First, we show in the third row of \F\,\ref{fig:stab_k16} and in \F\,\ref{fig:stab_krest_single} stability maps for integrations in which the stellar binary was effectively replaced by a point mass. For Kepler 16, the stable regions are significantly larger in the single-star case, especially for retrograde orbits. This shows that the binary tends to destabilize the moon, as is intuitively clear. 

Before discussing mean motion commensurabilities (MMCs), we note that a possible explanation for the differences between the single- and binary-star cases is that the binary could impose precession of the angular-momentum vector of the orbit of the planet (i.e., nodal precession, or a non-zero $\dot{\Omega}_\p$), which could affect the secular evolution of the planet-moon pair \citep{2015MNRAS.449.4221H,2015PNAS..112.9264M,2017MNRAS.470.1657H,2018MNRAS.474.3547G}. This can be tested by comparing the LK time-scale of the planet-moon pair, $t_\mathrm{LK}$, to the time-scales of nodal precession induced by the binary, $t_{\mathrm{prec},\,\bin}$. Specifically, we consider the ratio (\citealt{2017MNRAS.470.1657H}, eq. 32) 
\begin{align}
\beta \equiv \frac{t_\mathrm{LK}}{t_{\mathrm{prec},\,\bin}} \simeq \frac{3}{4} \left ( \frac{a_\bin}{a_\m} \right )^{3/2} \left ( \frac{M_\p}{M_1+M_2} \right )^{3/2},
\end{align}
where we set the mutual inclination between the orbits of the stellar binary and the CBP to be zero, as appropriate for the {\it Kepler} CBP systems. If $\beta\ll 1$, then the nodal precession by the binary on the CBP orbit is slow compared to the LK evolution of the planet-moon pair, and in that sense the `binarity' of the stellar binary is unimportant. Evaluating $\beta$ for the {\it Kepler} systems and the ranges of $a_\m$ considered in our simulations, we find that $\beta$ is typically small, with a mean value of $\simeq 1\times 10^{-4}$, and a maximum value of $\simeq 2 \times 10^{-3}$. We conclude that nodal precession on the orbit of the CBP around the stellar binary can be neglected in terms of the secular planet-moon evolution.

Therefore, we consider an alternative explanation based on MMCs. Commensurabilities of mean motions are known to lead to mean motion {\it resonances} (MMRs). The latter are associated with oscillations of the orbital elements, with important implications in a large number of contexts, including multiplanet systems (e.g., \citealt{1976ARA&A..14..215P,1984CeMec..32..307S,2013A&A...556A..28B,2013ApJ...774..129D}), the Kirkwood gaps in the Asteroid belt (e.g., \citealt{1983Icar...56...51W,1990CeMDA..47...99H,1997AJ....114.1246M}), the rings of Saturn, e.g., \citealt{1978Icar...34..240G,1982Natur.299..209B}), and binary stars (e.g., \citealt{1999AJ....117..621H,2018AJ....155..174T}). MMRs can be associated with stabilizing effects (e.g., the Galilean satellites), or destabilizing in the sense that a strong resonance can drive large orbital variations, ultimately leading to dynamical instability. The dynamics of MMRs in quadruple systems in the 2+2 configuration, which applies to our systems, are not well-understood (see \citealt{2018MNRAS.475.5215B} for a pioneering study which applies to the case of four bodies with comparable masses). 

In our case, it is reasonable to expect that there could be a MMR of the stellar binary with the orbit of the moon around the planet if the mean motions are commensurate, and that the MMR could have a destabilizing effect if the strength of the resonance is large enough (i.e., large enough to cause significant variations in the semimajor axis of the moon, triggering dynamical instabilities). 

Using Kepler's law, it is straightforward to show that the condition $P_\bin = \alpha P_\m$, where $P_\bin$ and $P_\m$ are the binary and lunar orbital periods, respectively, and $\alpha$ is dimensionless, can be written as
\begin{align}
\label{eq:a_MMC}
a_{\m,\,\MMC} = \alpha^{-2/3} \, a_\bin \left(\frac{M_\p}{M_1+M_2} \right )^{1/3}.
\end{align}
This expression has the same dependence on the masses as the Hill radius (equation~\ref{eq:r_H}), but the dependence on $a_\p(1-e_\p)$ is replaced by $a_\bin$. For the 1:1 MMC, $\alpha=1$; the corresponding values of $a_{\m,\,\MMC}$ are shown in the stability maps with the solid black lines (the 2:1 and 1:2 commensurabilities are shown with the black dashed lines). 

As shown in the stability maps, the 1:1 MMC expression captures the boundary between stable and unstable orbits well for most systems, especially for Kepler 35, 38, 47b, 64, and 413, which do not show a strong dependence on $i_\pm$ ignoring the region of LK-induced collisions near high inclinations. There are two notable exceptions: for Kepler 47c and Kepler 1647, the stability boundary is significantly larger than $a_{\m,\,\MMC}$, and is more consistent with $r_\mathrm{H}$. In the latter two systems, the CBP semimajor axis is relatively large (more than several times the critical separation for dynamical stability, see, e.g., fig. 1 of \citealt{2018ApJ...858...86F}), such that the effect of the 1:1 MMC is expected to be weak, and should not set the stability boundary. This is supported by a comparison of the maps for Kepler 47c and 1647 in Figs\,\ref{fig:stab_krest} and \ref{fig:stab_krest_single}, which reveals that the stability regions for these systems are virtually identical for the single and binary star cases.

\subsection{Collisions and strong interactions}
\label{sect:results:int}
As shown in the stability maps, most collisions occur at high inclinations around $90^\circ$. In Table\,\ref{table:col}, we show the number of collisions of the moon with the planet ($N_\mathrm{col,\,\p}$), the primary star ($N_\mathrm{col,\,\star1}$), and the secondary star ($N_\mathrm{col,\,\star2}$, if applicable). Note that the total number of integrations for each {\it Kepler} system is 16,000. The majority of the collisions are between the moon with the planet. Collisions of the moon with the stars are rare. 

\begin{table*}
\begin{tabular}{lcccccccc}
\toprule
& \multicolumn{3}{c}{Binary star ($q_\m=0.001$)} & \multicolumn{3}{c}{Binary star ($q_\m=0.1$)} & \multicolumn{2}{c}{Single star ($q_\m=0.001$)} \\
& $N_\mathrm{col,\,\p}$ & $N_\mathrm{col,\,\star1}$ & $N_\mathrm{col,\,\star2}$ & $N_\mathrm{col,\,\p}$ & $N_\mathrm{col,\,\star1}$ & $N_\mathrm{col,\,\star2}$ & $N_\mathrm{col,\,\p}$ & $N_\mathrm{col,\,\star}$ \\
\midrule
Kepler 16  &  3235  &  106  &  129 &  3366  &  79  &  110 &  3193  &  0  \\
Kepler 34  &  3120  &  3  &  3 &  3161  &  1  &  2 &  3395  &  0   \\
Kepler 35  &  2589  &  26  &  22 &  2695  &  22  &  23 &  4376  &  0  \\
Kepler 38  &  5743  &  0  &  0 &  5690  &  0  &  0 &  6059  &  0  \\
Kepler 47b  &  4056  &  0  &  0 &  4246  &  0  &  0 &  4205  &  0   \\
Kepler 47c  &  3115  &  0  &  0 &  3112  &  0  &  0 &  3281  &  0  \\
Kepler 64  &  2442  &  55  &  73 &  2648  &  55  &  87 &  3319  &  1 \\
Kepler 413  &  3155  &  15  &  10 &  3246  &  11  &  10 &  3958  &  0  \\
Kepler 453  &  3047  &  0  &  0 &  3098  &  0  &  0 &  4727  &  0  \\
Kepler 1647  &  1660  &  0  &  0 &  1702  &  0  &  0 &  1977  &  0 \\
\bottomrule
\end{tabular}
\caption{ Number of collisions of the moon with the planet ($N_\mathrm{col,\,\p}$), the primary star ($N_\mathrm{col,\,\star1}$), and the secondary star ($N_\mathrm{col,\,\star2}$, if applicable), in the simulations with $q_\m=0.001$ (binary star case), $q_\m=0.1$ (binary star case), and $q_\m=0.001$ (single star case), and for an integration time of $1000\,P_\p$. The total number of integrations for each {\it Kepler} system is 16,000. }
\label{table:col}
\end{table*}

The high-inclination collision boundary has a `funnel' shape which is symmetric around $90^\circ$, and becomes wider at smaller $a_\m$. This can be understood from standard LK dynamics \citep{1962P&SS....9..719L,1962AJ.....67..591K}. In particular, in the quadrupole-order test-particle limit and treating the binary as a point mass, the maximum eccentricity for zero initial eccentricity reached is given by
\begin{align}
\label{eq:e_max_LK}
e_{\m,\,\mathrm{max}} = \sqrt{1-\frac{5}{3} \cos^2(i_\pm)}.
\end{align}
The implied periapsis distance is $r_{\peri,\,\m} = a_\m (1-e_{\m,\,\mathrm{max}})$ for $\cos^2(i_\pm) < 3/5$. Equating $r_{\peri,\,\m}$ to the planetary radius $R_\p$ gives a relation between $a_\m$ and $i_\pm$, which is shown in the stability maps with the red dashed lines. These lines generally agree with the simulations at small $a_\m$, at which collisions are driven by LK evolution. At larger $a_\m$, collisions are due to short-term dynamical instabilities rather than secular evolution, and become less dependent on inclination. 

The stability maps shown above are based on Newtonian point-mass dynamics, neglecting strong interactions such as tidal effects and tidal disruption. Here, we briefly consider such interactions when the moon passes close to its parent planet (we do not consider strong interactions with the stars).

\begin{figure}
\center
\includegraphics[scale = 0.47, trim = 5mm 0mm 0mm 0mm]{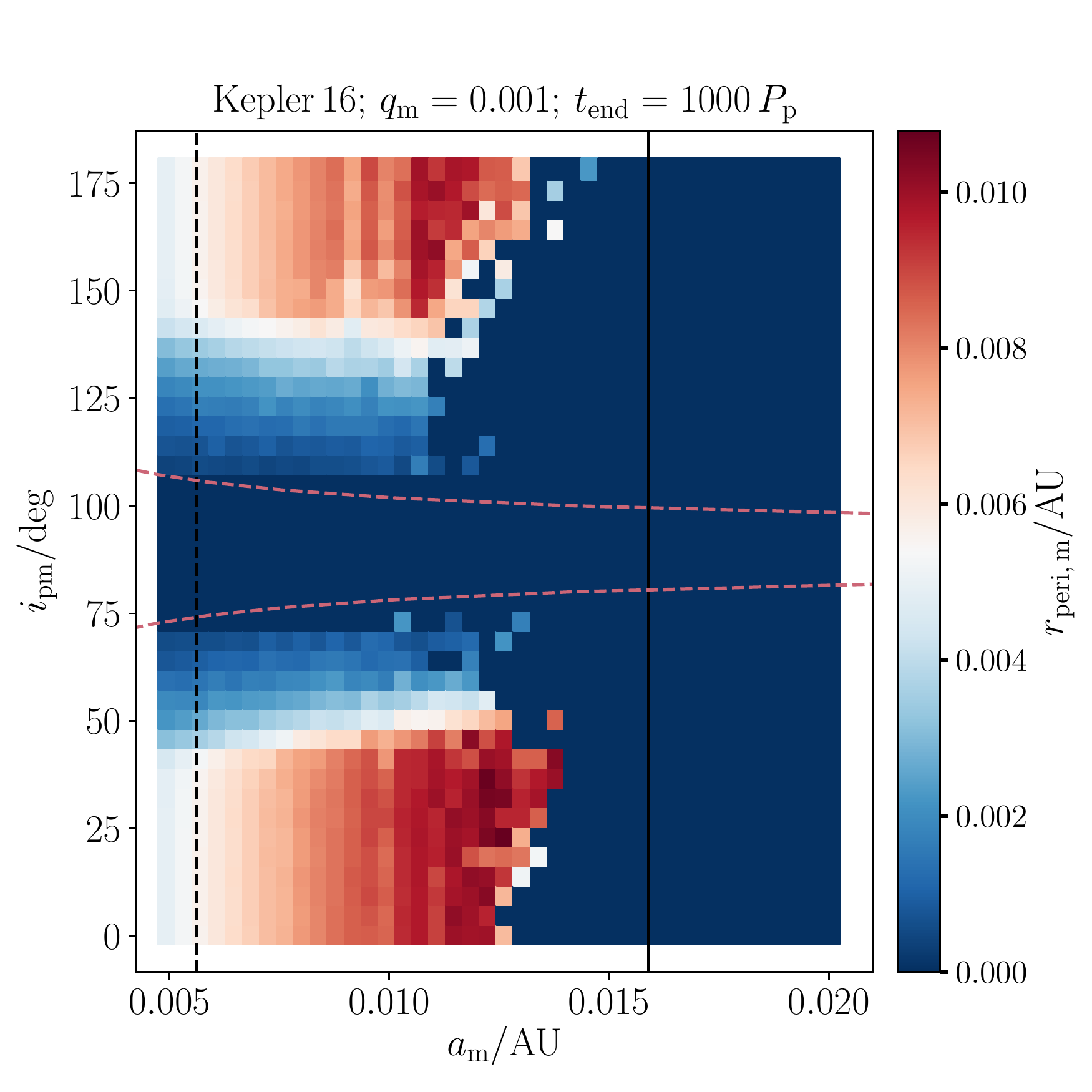}
\caption{\small Minimum periapsis distances, $r_{\peri,\,\m}$, recorded in the simulations in the $(a_\m,i_\pm)$ plane for Kepler 16. For unbound orbits or collisions, we set $r_{\peri,\,\m}$ to $-0.001\,\au$, which is indicated with the dark blue regions in the right-hand regions of each panel. The vertical black solid lines show the location of the 1:1 MMC of the moon with the binary (equation~\ref{eq:a_MMC}), and the vertical black dashed line shows the locations of the 2:1 MMC. The red dashed lines show the boundary for collisions of the moon with the planet due to LK evolution, computed according to equation~(\ref{eq:e_max_LK}).}
\label{fig:rps_k16}
\end{figure}

We evaluate the importance of strong interactions by comparing the periapsis distances of the moon to its parent planet to (multiples of) the radius of the planet. In \F\,\ref{fig:rps_k16}, we show for Kepler 16 in the $(a_\m,i_\pm)$ plane the minimum periapsis distances, $r_{\peri,\,\m}$, recorded in the simulations. Here, we determine $r_{\peri,\,\m} = a_\m(1-e_\m)$ from the minimum value among the systems in the $(a_\m,i_\pm)$ plane that are stable (i.e., if the moon remains bound to the planet for all 10 realizations of $\mathcal{M}_\m$); if at least one of the realizations yielded an unbound orbit or a collision, then we set $r_{\peri,\,\m}$ to $-0.001\,\au$, which is indicated in \F\,\ref{fig:rps_k16} with the dark blue regions. Note that, due to a finite number of output snapshots, the true closest approach can be missed in some cases. For low inclinations, there is no excitation of $e_\m$, and $r_{\peri,\,\m} = a_\m(1-e_{\m,\,0})$, where $e_{\m,\,0}$ is the initial eccentricity. This is manifested in \F\,\ref{fig:rps_k16} as a linear relation for small inclinations. For larger inclinations, LK evolution drives high $e_\m$ and small $r_{\peri,\,\m}$, and collisions with the planet occur as $i_\pm$ approaches $90^\circ$.

\begin{figure}
\center
\includegraphics[scale = 0.45, trim = 0mm 0mm 0mm 0mm]{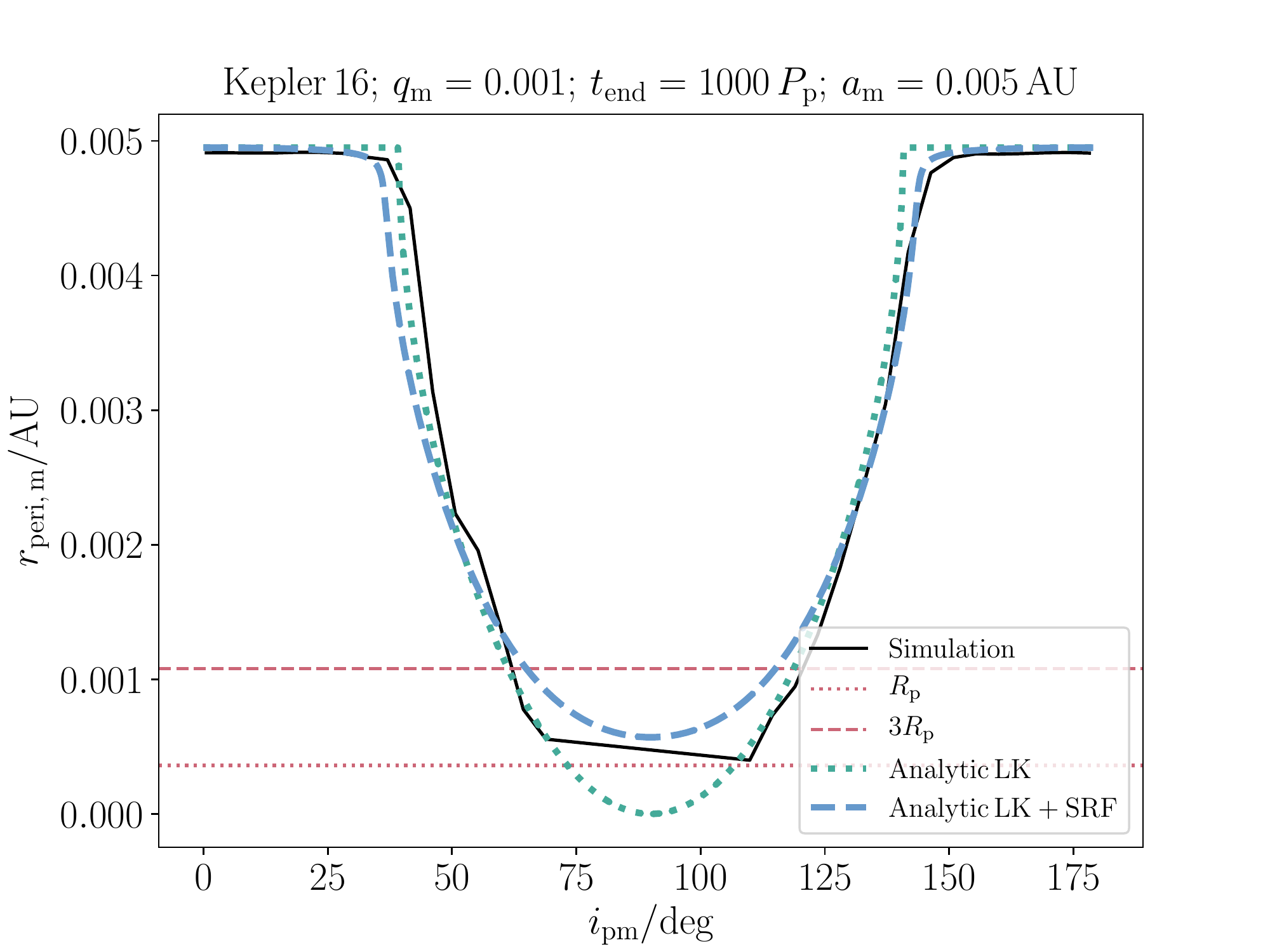}
\caption{\small Black solid lines: minimum periapsis distances $r_{\peri,\,\m}$ for Kepler 16 as in \F\,\ref{fig:rps_k16}, here for a slice in semimajor axis, i.e., $a_\m = 0.005 \, \au$. Green dotted curve: the analytic LK periapsis distance based on equation~(\ref{eq:e_max_LK}). Blue dashed curve: analytic LK periapsis distance based on a semianalytic calculation taking into account the oblateness due to rotation and tidal bulges of the planet (assuming a spin rotation period of 5 hr  and an apsidal motion constant of 0.25), and relativistic precession (which is not important). The red dotted and dashed lines show $r_{\peri,\,\m} = R_\p$ and $r_{\peri,\,\m} = 3\, R_\p$, respectively, where $R_\p = 0.754\,R_\mathrm{J}$ is the planetary radius \citep{2011Sci...333.1602D}. }
\label{fig:rps_slice_k16}
\end{figure}

Also, we show in \F\,\ref{fig:rps_slice_k16} a plot of $r_{\peri,\,\m}$ versus $i_\pm$ for a slice in $a_\m$, with $a_\m = 0.005 \, \au$. The latter figure shows that the dependence of $r_{\peri,\,\m}$ on $i_\pm$ for $r_{\peri,\,\m}>R_\p$ is well described by the canonical LK relation (equation~\ref{eq:e_max_LK}; green dashed line). The red dotted and dashed lines show $r_{\peri,\,\m} = R_\p$ and $r_{\peri,\,\m} = 3\, R_\p$, respectively, where $R_\p = 0.754\,R_\mathrm{J}$ is the planetary radius \citep{2011Sci...333.1602D}. Here, we take a factor of 3 to be indicative of the regime where tidal effects are important. Much closer in, tidal disruption is possible, or even direct collision. In the case shown for Kepler 16 and for $a_\m=0.005\,\au$, tidal effects are potentially important for inclinations near $60^\circ$ and $120^\circ$. Direct collision occur for inclinations between $\sim 75^\circ$ and $105^\circ$.

\subsection{Short-range forces}
\label{sect:results:srf}
Our $N$-body integrations include the Newtonian point-mass terms only. Here, we investigate, a posteriori, the importance of short-range forces (SRFs) acting on the planet-moon orbit, in conjunction with LK cycles. Such SRFs can include relativistic precession, and precession due to the oblateness of the planet. In the case of Kepler 16, relativistic precession is completely unimportant since the precession time-scale (on the order of Myr) is much longer than the LK time-scale (on the order of tens of years; see also Table\,\ref{table:SRF}). 

However, the oblateness of the planet can be important depending on the assumed structure of the planet, and, more importantly, its rotation rate. In \F\,\ref{fig:rps_slice_k16}, we show with the blue dashed line the periapsis distance $r_{\peri,\,\m}$ according to LK theory with the addition of the oblateness due to rotation and tidal bulges of the planet (assuming $q_\m=0.001$). We compute $e_{\m,\,\mathrm{max}}$ by using energy and angular-momentum conservation, with the expressions for the Hamiltonian adopted from \citet{2007ApJ...669.1298F}. Here, we use the fact that the maximum eccentricity occurs when the inner orbit argument of periapsis is $\omega_\m=90^\circ$, and assumed $L_\p\gg L_\m$ in the expression for conservation of angular momentum (where $L$ is the orbital angular momentum), such that $\cos(i_\pm)\sqrt{1-e_\m^2}$ is constant (this method is equivalent to that of \citealt{2015MNRAS.447..747L}). For the planet, we here assume a spin rotation period of 5 hr (a somewhat extreme case, given that for Kepler 16 the critical rotation period for centrifugal breakup is $2\pi/\sqrt{GM_\p/R_\p^3} \simeq 3.4\,\mathrm{hr}$), and an apsidal motion constant of 0.25. For these values, there is some deviation from the canonical relation equation~(\ref{eq:e_max_LK}), although, in this case, $r_{\peri,\,\m}$ is not much affected unless $i_\pm$ is close to $90^\circ$. 

\begin{table}
\scriptsize
\begin{tabular}{lcccccccc}
\toprule
& \multicolumn{2}{c}{$t_\mathrm{LK}/\mathrm{yr}$} &\multicolumn{2}{c}{$t_\mathrm{1PN}/\mathrm{Myr}$} & \multicolumn{2}{c}{$t_\mathrm{TB}/\mathrm{Myr}$} & \multicolumn{2}{c}{$t_\mathrm{rot}/\mathrm{yr}$} \\
& $a_0$ & $a_1$& $a_0$ & $a_1$& $a_0$ & $a_1$& $a_0$ & $a_1$ \\
\midrule
K16		& 19.8  &  3.5  &  5.3  &  95.0  &  3.5  &  6400  &  9.7  &  557.5  \\
K34		& 24.4  &  7.8  &  9.8  &  65.3  &  4.0  &  551.9  &  11.6  &  165.5  \\
K35		& 4.0  &  2.3  &  22.4  &  57.8  &  6.7  &  78.9  &  16.9  &  63.7  \\
K38		& 3.3  &  1.6  &  1061  &  3520  &  0.001  &  0.028  &  0.46  &  2.5  \\
K47b		& 8.3  &  0.9  &  0.7  &  29.8  &  0.001  &  10.6  &  0.09  &  15.5  \\
K47c		& 138  &  25.6  &  0.9  &  15.2  &  0.004  &  6.5  &  0.25  &  12.7  \\
K64		& 9.0  &  2.6  &  2.6  &  21.0  &  13.2  &  2985  &  14.4  &  267.1  \\
K413		& 14.5  &  1.2  &  0.2  &  12.5  &  0.003  &  192.0  &  0.17  &  59.2  \\
K453		& 10.6  &  5.8  &  1147  &  3149  &  0.011  &  0.15  &  1.49  &  6.1  \\
K1647	& 349  &  297  &  3.1  &  4.0  &  27.0  &  53.9  &  26.1  &  37.9  \\
\bottomrule
\end{tabular}
\caption{ Short-range precession time-scales for lunar orbits around the {\it Kepler} CBPs. The time-scales given are the LK, relativistic, TB, and rotation time-scales (see text for details), and are shown for two semimajor axes: $a_\m = a_0$, the smallest semimajor axis considered in the simulations, and $a_\m = a_1 = a_\mathrm{\m,\,\MMC}$, the location of the 1:1 MMC with the binary. Note that the units for $t_\mathrm{LK}$ and $t_\mathrm{rot}$ are yr, and Myr for $t_\mathrm{1PN}$ and $t_\mathrm{TB}$. }
\label{table:SRF}
\end{table}

To investigate the importance of SRF in all {\it Kepler} CBP systems, we compute the associated time-scales in Table\,\ref{table:SRF}. Approximating the binary as a point mass, the LK time-scale can be estimated as (e.g., \citealt{1997AJ....113.1915I,1999CeMDA..75..125K,2015MNRAS.452.3610A})
\begin{align}
\label{eq_t_LK}
t_\mathrm{LK} &= \frac{P_\p^2}{P_\m} \frac{M_\p + M_\m + M_\bin}{M_\bin} \left (1-e_\p^2 \right )^{3/2},
\end{align}
where $M_\bin \equiv M_1+M_2$. The relativistic time-scale is (e.g., \citealt{1972gcpa.book.....W})
\begin{align}
\label{eq:t_PN}
t_\mathrm{1PN} = \frac{1}{6} \frac{a_\m c^2}{GM_\p} P_\m,
\end{align}
and the tidal bulges (TB) and rotation time-scales are given by (e.g., \citealt{2007ApJ...669.1298F})
\begin{align}
\label{eq:t_TB}
t_\mathrm{TB}^{-1} &= \frac{15}{8} n_\m \frac{M_\m}{M_\p} k_\mathrm{AM,\,p} \left ( \frac{R_\p}{a_\m} \right )^5; \\
\label{eq_t_rot}
t_\mathrm{rot}^{-1} &=  n_\m \left (1 + \frac{M_\m}{M_\p} \right ) k_\mathrm{AM,\,p} \left ( \frac{R_\p}{a_\m} \right )^5 \left ( \frac{\Omega_\p}{n_\m} \right )^2,
\end{align}
where $n_\m = 2\pi/P_\m$ is the lunar mean motion, and $k_\mathrm{AM,\,\p}$ is the planetary apsidal motion constant. In equations~(\ref{eq:t_PN}), (\ref{eq:t_TB}) and (\ref{eq_t_rot}), we have set the eccentricity of the lunar orbit to zero. 

We remark that the regime in which the LK and rotation time-scales are comparable is characterized by the Laplace radius, $a_{\m,\,\mathrm{L}}$ (\citealt{2009AJ....137.3706T}; by equating equation~\ref{eq_t_LK} and the inverse of equation~\ref{eq_t_rot}, one can obtain a relation for the Laplace radius, $a_{\m,\,\mathrm{L}}$, which is equivalent to eq. 24 of the latter paper except for a constant factor). If $a_\m \ll a_{\m,\,\mathrm{L}}$, then the dynamics of the moon are dominated by the oblateness of the planet due to rotation, whereas the latter can be neglected if $a_\m \gg a_{\m,\,\mathrm{L}}$. 

In Table\,\ref{table:SRF}, we adopt the measured planetary radii (see Table\,\ref{table:IC}), set $q_\m = 0.001$ and $k_\mathrm{AM,\,\p}=0.25$, and we assume that the planet is spinning at half its breakup rotation speed, i.e., $\Omega_\p = (1/2) \,\Omega_{\p,\,\mathrm{crit}} = (1/2) \sqrt{GM_\p/R_\p^3}$. The time-scales are shown for two semimajor axes: $a_\m = a_0$, the smallest semimajor axis considered in the simulations, and $a_\m = a_1 = a_\mathrm{\m,\,MM}$, the location of the 1:1 MMC with the binary. 

As shown in Table\,\ref{table:SRF}, relativistic precession and tidal bulges are completely negligible for the {\it Kepler} CBP systems. Precession due to oblateness induced by rotation of the planet is potentially important at small semimajor axes, especially for Kepler 38, 47b, 47c, 413, 453, and 1647. At larger semimajor axes ($a_\m=a_{\m,\,\MMC}$), precession due to rotation is potentially important for Kepler 38, 47c, 453 and 1647. However, it should be taken into account that we assumed a somewhat extreme case of a near-critical rotation speed of the planet, and that $t_\mathrm{rot} \propto \Omega_\p^{-2}$. For example, for Kepler 38, 47c and 453, precession due to rotation would no longer be dominant at $a_\m=a_{\m,\,\MMC}$ if $\Omega_\p = (1/4) \,\Omega_{\p,\,\mathrm{crit}}$ (an increase in $t_\mathrm{rot}$ by a factor of four).

\subsection{Summary plots}
\label{sect:results:sum}
\begin{figure}
\center
\includegraphics[scale = 0.45, trim = 0mm 0mm 0mm 0mm]{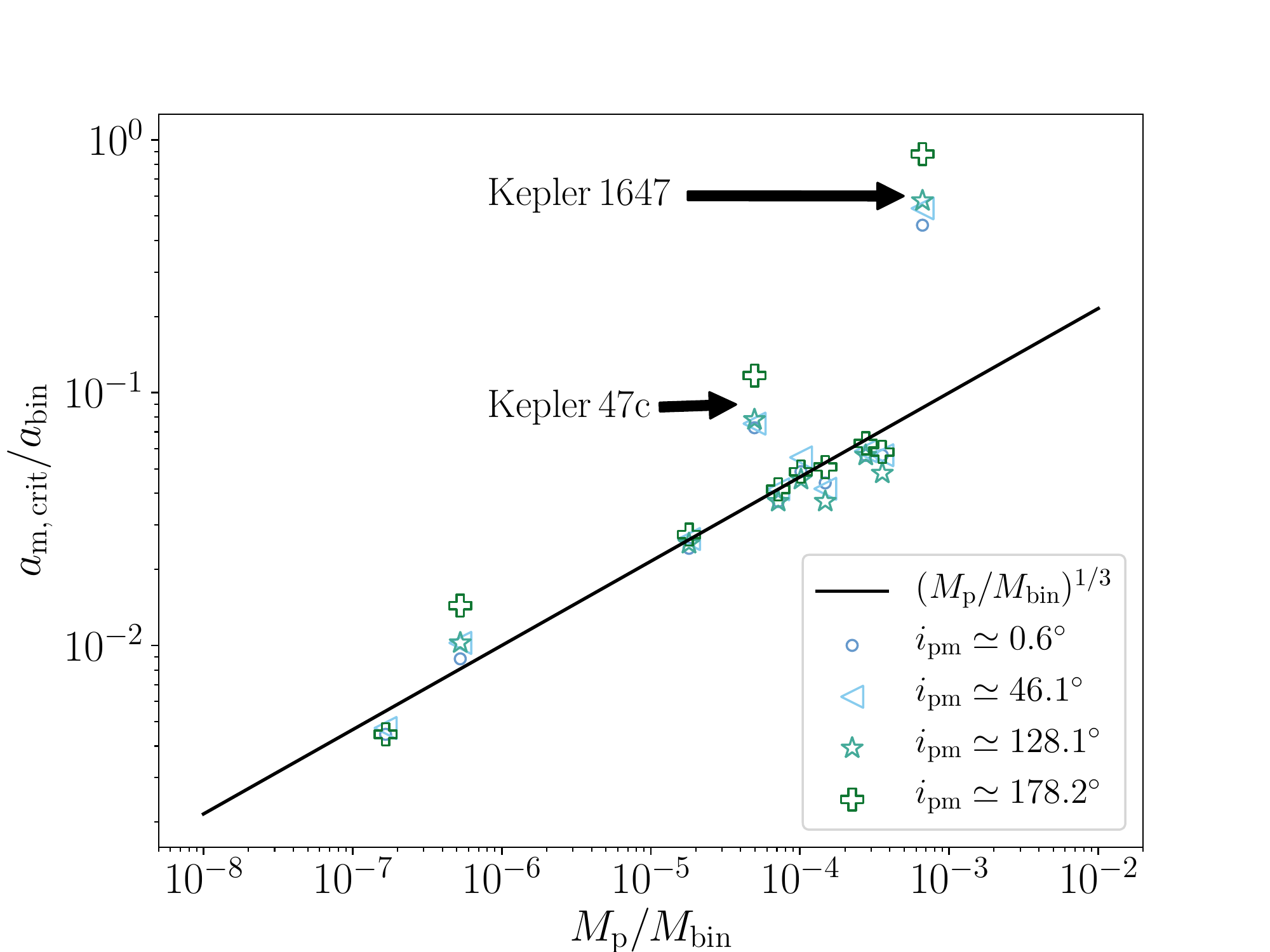}
\includegraphics[scale = 0.45, trim = 0mm 10mm 0mm 0mm]{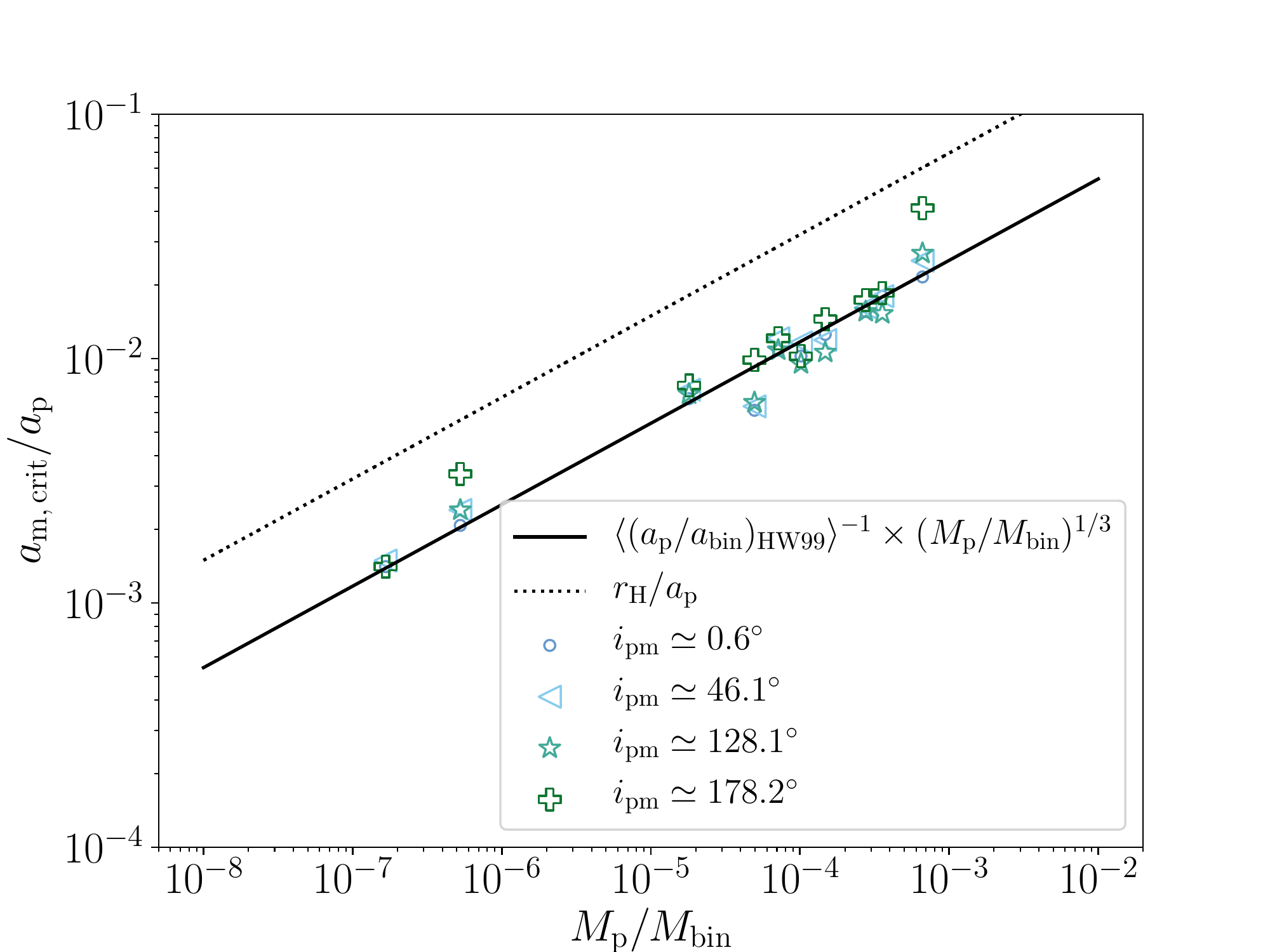}
\caption{\small Top panel: largest stable value of $a_\m$, $a_{\m,\,\crit}$, normalized to $a_\bin$ and plotted as a function of $M_\p/M_\bin$, where $M_\bin = M_1+M_2$, for each system in the binary case with $q_\m = 0.001$, and for a given value of $i_\pm$ (refer to the legend). The solid black line shows equation~(\ref{eq:a_MMC}), the location of the 1:1 MMC. Bottom panel: $a_{\m,\,\crit}$ normalized to $a_\p$. The black dotted line shows the Hill radius (equation~\ref{eq:r_H}), and the black solid line shows the estimate equation~(\ref{eq:a_m_a_p_HW99}) assuming the CBP is close to the critical boundary for stability. }
\label{fig:summary}
\end{figure}

We summarize the results of the stability maps by determining the largest stable values of $a_\m$, $a_{\m,\,\crit}$, for each system and for a given value of $i_\pm$, and plotting these values as a function of the masses of the planets and the stars (assuming $q_\m = 0.001$). In particular, we show in the top panel of \F\,\ref{fig:summary} $a_{\m,\,\crit}$ normalized to $a_\bin$ as a function of $M_\p/M_\bin$. Different symbols correspond to different inclinations (refer to the legend). Note that high inclinations, near $90^\circ$, are excluded since in this case there are no stable orbits in our simulations. Also plotted in the same panel with the solid black line is equation~(\ref{eq:a_MMC}), the location of the 1:1 MMC. Most of the data points from the simulations match well with equation~(\ref{eq:a_MMC}), again showing that the critical semimajor axis is close to the 1:1 MMC. Two notable exceptions are Kepler 47c and Kepler 1647 (indicated in the figure with arrows), which, as mentioned in \Se\,\ref{sect:results:MMC}, have relatively large planetary semimajor axes, such that the effect of the 1:1 MMC is expected to be weak and therefore it would not set the stability boundary.

Alternatively, one can normalize $a_{\m,\,\crit}$ by $a_\p$. The resulting points from the simulations are shown in the bottom panel of \F\,\ref{fig:summary}. Again, most of the data are well fit by a power-law function of $M_\p/M_\bin$ with a slope of $1/3$, although the absolute normalization is different. The black dashed line shows the Hill radius (equation~\ref{eq:r_H}), which, as noted above, overestimates the critical $a_\m$ for stability. However, we can obtain a relation that better describes the data by using the fact that most of the {\it Kepler} CBPs are close to the limit for stability (i.e., stability of the planet in absence of the moon). In particular, 
\begin{align}
\frac{a_\m}{a_\p} = \frac{a_\m}{a_\bin} \frac{a_\bin}{a_\p} = \left ( \frac{M_\p}{M_\bin} \right )^{1/3} \frac{a_\bin}{a_\p},
\end{align}
where we used equation~(\ref{eq:a_MMC}) with $\alpha=1$ (the 1:1 MMC). Subsequently, we replace $a_\bin/a_\p$ by  $\langle (a_\p/a_\bin)_\mathrm{HW99} \rangle^{-1}$, where $(a_\p/a_\bin)_\mathrm{HW99}$ is the critical CBP semimajor axis in units of $a_\bin$ determined from the analytic fits of \citet{1999AJ....117..621H}, and where the average is taken assuming a thermal eccentricity distribution of $e_\bin$ and a flat mass ratio distribution (i.e., flat in $q = M_2/M_1$). With these assumptions, we find $\langle (a_\p/a_\bin)_\mathrm{HW99} \rangle^{-1} \simeq 0.252267$, such that\footnote{Alternatively, one could compute $\langle 1/(a_\p/a_\bin)_\mathrm{HW99} \rangle \simeq 0.25592$, which gives nearly the same result.}
\begin{align}
\label{eq:a_m_a_p_HW99}
\frac{a_\m}{a_\p} \simeq 0.252267 \, \left ( \frac{M_\p}{M_\bin} \right )^{1/3}.
\end{align}
Equation~(\ref{eq:a_m_a_p_HW99}) is shown in the bottom panel of \F\,\ref{fig:summary} with the black solid line, and agrees with most of the data from the simulations.

\section{Discussion}
\label{sect:discussion}

\subsection{Caveats related to the {\it Kepler} systems}
Kepler 47 harbors at least two CBPs, Kepler 47b and Kepler 47c \citep{2012Sci...337.1511O}, and there may be a third CBP, Kepler 47d \citep{2015ApJ...799...88H}. In our work, we did not consider Kepler 47d, which has not yet been confirmed. Also, we treated the two confirmed planets in Kepler 47 as separate cases, i.e., in each case we neglected the gravitational potential of the other CBP. We have checked that this assumption is justified by running simulations with both confirmed planets included, which yielded no significantly different results. 

Kepler 64, also known as Planet Hunters 1 (PH1), is a quadruple-star system; the binary+CBP system is orbited by a distant stellar binary (a visual binary) at a separation of $~\sim 1000\,\au$ \citep{2013ApJ...768..127S}. In our integrations, we did not include the visual binary. This is justified, given the large separation between the two stellar binaries, and the relative scale of the inner CBP+planet triplet.

\subsection{The role of mean motion resonances}
We have shown that, with the exception of two {\it Kepler} systems (Kepler 47c and 1647), there is good agreement of the locations of the 1:1 commensurability lines of the moon with the stellar binary (equation~\ref{eq:a_MMC}) with the empirical stability boundaries, for both prograde and retrograde orbits. However, this does not provide proof that the instability is triggered by mean motion {\it resonances}. An investigation into the latter possibility is beyond the scope of this work, but merits future attention.

\subsection{Implications of collisions}
\label{sect:discussion:col}
As shown above, most of the collisions in our simulations are collisions of the moon with the CBP, and occur at high inclination. However, some of the collisions are between the moon and the stars, and show no strong inclination dependence. Such collisions could enhance the metallicity of the stars, and/or trigger a speed-up of the stellar rotation in the case of a relative massive moon and a low-mass star. Speed-up of stellar rotation has been proposed to occur as a result of planets plunging onto their host star due to high-eccentricity migration \citep{2018arXiv180208260Q}.

\section{Conclusions}
\label{sect:conclusions}
We investigated the stability of moons around planets in stellar binary systems. In particular, we considered exomoons around the transiting {\it Kepler} circumbinary planets (CBPs). Such exomoons may be suitable for harboring life, and are potentially detectable in future observations. We carried out numerical $N$-body simulations and determined regions of stability around the {\it Kepler} CBPs. Our main conclusions are listed below.

\medskip \noindent 1. We obtained stability maps in the $(a_\m,i_\pm)$ plane, where $a_\m$ is the lunar semimajor axis with respect to the CBP, and $i_\pm$ is the inclination of the orbit of the moon around the planet with respect to the orbit of the planet around the stellar binary. For most of the {\it Kepler} CBPs and ignoring the dependence on $i_\pm$, the stability regions are well described by the location of the 1:1 mean motion commensurability (MMC) of the binary orbit with the orbit of the moon around the CBP (equation~\ref{eq:a_MMC}). This is related to a destabilizing effect of the binary compared to the case if the binary were replaced by a single body, and which is borne out by corresponding 3-body integrations. For the two exceptions, Kepler 47b and Kepler 1647, the CBP semimajor axis is relatively large such that the effect of the 1:1 MMC is expected to be weak and therefore it would not set the stability boundary.

\medskip \noindent 2. For stable lunar orbits and high inclinations, $i_\pm$ near $90^\circ$, the evolution is dominated by Lidov-Kozai oscillations \citep{1962P&SS....9..719L,1962AJ.....67..591K}. These imply that moons in orbits that are dynamically stable could be brought to close proximity within the CBP, and experience strong interactions such as tidal evolution, tidal disruption, or direct collisions. This suggests that there is a dearth of highly-inclined exomoons around the {\it Kepler} CBPs, whereas coplanar exomoons are dynamically allowed.

\medskip \noindent 3. Most of the collisions in our simulations are CBP-moon collisions, occurring at high inclination. Collisions with the stars are rare. However, if such collisions do occur, the stars in the binary might be enhanced in metallicity, and/or show an anomalously-high rotation speed.

\section*{Acknowledgements}
This paper is part of the {\it Moving Planets Around} educational book project, which is supported by Piet Hut, Jun Makino, and the RIKEN Center for Computational Science. We thank the referee, Evgeni Grishin, for a very helpful and detailed report. Also, we thank Scott Tremaine and Daniel Fabrycky for discussions, and Ren\'{e} Heller for comments on the manuscript. ASH gratefully acknowledges support from the Institute for Advanced Study, The Peter Svennilson Membership, and NASA grant NNX14AM24G. This work was partially supported by the Netherlands Research Council NWO (grants \#643.200.503, \#639.073.803 and \#614.061.608) by the Netherlands Research School for Astronomy (NOVA). This research was partially supported by the Interuniversity Attraction Poles Programme (initiated by the Belgian Science Policy Office, IAP P7/08 CHARM) and by the European Union's Horizon 2020 research and innovation programme under grant agreement No 671564 (COMPAT project). MXC acknowledges the support by the Institute for Advanced Study during his visit. Part of this research was carried out at the Jet Propulsion Laboratory, California Institute of Technology, under a contract with the National Aeronautics and Space Administration.

\bibliographystyle{mnras}
\bibliography{literature}

\label{lastpage}
\end{document}